\documentclass[aps,prb,twocolumn,groupedaddress,floatfix,longbibliography,superscriptaddress]{revtex4-2}

\usepackage{amsmath}
\usepackage{amssymb}
\usepackage{amsfonts}
\usepackage{bbold}
\usepackage{bm}
\usepackage{cancel}
\usepackage{times,float}
\usepackage{graphicx}
\usepackage[usenames,dvipsnames,svgnames]{xcolor}
\usepackage{hyperref}
\hypersetup{colorlinks=true, linkcolor=Blue, citecolor=Blue,urlcolor=Blue}
\usepackage{multirow}
\usepackage{ulem}
\usepackage{color,epsfig}
\usepackage{slashed}
\usepackage{feynmf}
\usepackage{array}
\usepackage{physics}
\usepackage{subcaption}
\usepackage{soul}
\usepackage{bm}
\usepackage{graphicx}
\graphicspath{{FIGURES/}}
\usepackage{bbold}
\usepackage{cancel}
\usepackage{mathrsfs}
\usepackage[compatibility=false]{caption}
\usepackage{times,float}
\usepackage{physics}
\usepackage{caption}
\usepackage{xcolor}
\usepackage{subcaption}

\begin{document}

\title{Steering of topological criticality in the Su-Schrieffer-Heeger chain via non-Hermitian substitutional impurities}

\author{Juan A. Cañas}
\email{juan.canas@correo.nucleares.unam.mx}
\affiliation{Instituto de Ciencias Nucleares, Universidad Nacional Autónoma de México, 04510 Ciudad de México, México}
\author{Guo-Qing Zhang}
\affiliation{Research Center for Quantum Physics, Huzhou Normal University, Huzhou 313000, People's Republic of China}
\author{A. Martín-Ruiz}
\affiliation{Instituto de Ciencias Nucleares, Universidad Nacional Autónoma de México, 04510 Ciudad de México, México}

\begin{abstract}
We investigate the topological phase in a one-dimensional Su-Schrieffer-Heeger (SSH) chain in the presence of a finite concentration of non-Hermitian substitutional impurities. Our primary focus is the discovery of a deterministic mechanism to actively reconfigure the topological phase boundaries by introducing site-dependent, non-reciprocal intracell and intercell hopping amplitudes. Specifically, breaking the internal structural symmetry of the non-reciprocal impurity hoppings induces a targeted macroscopic displacement of the critical point. This mechanism allows for the controlled expansion or compression of the topological phase strictly via local defect asymmetry. We validate this boundary displacement through finite-size scaling analysis. The extracted correlation length critical exponents ($\nu \approx 1.455$ and $\nu \approx 1.397$ for symmetric and asymmetric configurations, respectively) confirm a disorder-induced departure from the pristine SSH universality class, with the exponent diverging at high defect densities. We also report an impurity-induced real-to-complex spectral transition geometrically characterized by the Non-Hermitian Skin Effect. Ultimately, our results establish how isolated non-Hermitian asymmetries cooperatively interact to deterministically shift macroscopic topological protection.
\end{abstract}

\maketitle 
%-----------------------------------------------------------
\section{Introduction}

The Su–Schrieffer–Heeger (SSH) model stands as one of the most paradigmatic and extensively studied frameworks in the field of one-dimensional topological insulators \cite{Asboth_2015, Hasan_RMP2010}. Originally introduced to describe the electronic properties of polyacetylene \cite{SSH_PRL1979}, the model captures the essential physics of a tight-binding chain with alternating nearest-neighbor hopping amplitudes. Despite its mathematical simplicity, the SSH model supports a non-trivial topological phase characterized by the emergence of zero-energy edge states localized at the boundaries of an open chain, which are robustly protected by chiral symmetry \cite{Bomantara_JPCM2026}. This remarkable property has established the SSH model as an ideal platform for investigating topological phases and their extensions, ranging from higher-dimensional generalizations \cite{Benalcazar_Sci2017, Liu_PRL2017} and enlarged unit cells \cite{Bomantara_JPCM2026} to the inclusion of long-range interactions \cite{Zhang2022, Perez_PRB2019}, multi-band structures \cite{LEE_ChJP2022} , and finite-size thermodynamic crossovers \cite{DaConceicao2026}.

The pristine SSH model assumes translational invariance and Hermiticity, features that are rarely realized in experimental settings. Real materials inevitably contain impurities, defects, and disorder, which break translational symmetry and can fundamentally alter the topological properties of the system. This has motivated a substantial body of work investigating the interplay between disorder and topology in SSH-type systems. A particularly striking phenomenon in this context is the topological Anderson insulator (TAI) phase \cite{Li_PRL2009}, wherein disorder, traditionally considered detrimental to topological order, can actually induce a non-trivial topological phase in an otherwise trivial system. TAI phases have been extensively studied in SSH models with various types of disorder, including uncorrelated random disorder, quasi-periodic modulations, and chiral disorder \cite{Mondragon_PRL2014,Meier_Sci2018,Wang_Cryst2026}. More recently, the extension of these ideas to non-Hermitian systems has opened a new frontier \cite{Bergholtz_RMP2021}. Non-Hermitian SSH models, realized through non-reciprocal hoppings or gain-loss mechanisms, exhibit a rich phenomenology including the non-Hermitian skin effect (NHSE) \cite{Yao_PRL2018,Zhang_AiP2022}. In the presence of disorder, non-Hermitian SSH chains undergo complex disorder-driven topological phase transitions, where non-Hermiticity can either stabilize or suppress the topological phase depending on the specific parameter regime \cite{Luo_APL2023,Luo_PRL2019}. Notably, the NHSE itself displays a high sensitivity to disorder; as the disorder strength increases, the macroscopic accumulation of boundary-localized skin modes is disrupted, causing the eigenstates to rapidly delocalize back into the bulk \cite{Liu_APhyS2022,Zhang_AiP2022}. This intricate interplay between disorder and non-Hermiticity is perhaps best exemplified by the non-Hermitian TAI. In this regime, the synergistic combination of both ingredients gives rise to robust topological phases that are fundamentally absent in purely Hermitian disordered systems and clean non-Hermitian frameworks alike \cite{Zhang_SciChP2020}.

Within this broader context, the introduction of non-Hermiticity in topological lattices has predominantly followed two distinct theoretical paradigms. The first paradigm treats non-Hermiticity and disorder as extensive, macroscopic properties of the system. In foundational works on non-Hermitian topological phases and TAIs \cite{Zhang_SciChP2020, Gong_PRX2018, Bergholtz_RMP2021}, non-reciprocity or complex potentials are introduced globally across the entire chain, often alongside widespread random disorder. This continuous distribution acts as an intrinsic structural modification of the entire bulk Hamiltonian. Conversely, an emerging second paradigm explores non-Hermiticity as a strictly localized phenomenon, typically induced through isolated impurities, point defects, or localized couplings to a host lattice \cite{Shi_PRB2026,Wang_En2025, Spring_SciPost2024, Hetenyi_PRB2025,Khan_JLTP2026,Wang_PRB2025}. In these localized setups, the non-Hermitian defect serves as an external probe or a highly sensitive scattering center. Such isolated impurities have been recently shown to drastically alter the spatial localization of modes, often drawing skin modes into the bulk or quenching boundary states, thereby challenging the conventional bulk-boundary correspondence without modifying the Hermitian nature of the underlying host lattice.

In this work, we adopt a fundamentally different approach. We consider a pristine, Hermitian SSH chain and introduce substitutional impurities that replace lattice sites and locally modify the hopping amplitudes in their vicinity. Physically, this scheme mirrors the replacement of a native atom in a crystal lattice, or its analogue in modern artificial quantum, photonic, and topoelectrical platforms \cite{Meier_Sci2018, Luo_APL2023, Pedraza2024}, with a distinct species, leading to an abrupt, localized variation in the transition amplitudes. Crucially, non-reciprocity is introduced only at the impurity site, through asymmetric hopping amplitudes for transitions towards and away from the defect. The remainder of the chain remains perfectly Hermitian and translationally invariant. This setup allows us to address a distinct physical question: How does a localized, non-Hermitian defect affect the topological properties of an otherwise clean and Hermitian SSH chain? Unlike the global non-Hermiticity studied by Zhang \textit{et al.} \cite{Zhang_SciChP2020}, our non-reciprocal perturbation is spatially confined; unlike the external impurity coupled to a non-reciprocal chain studied in recent works \cite{Wang_En2025, Wang_PRB2025}, our impurity is substitutional and the host remains Hermitian. Furthermore, while recent studies have explored Hermitian SSH chains featuring deterministically placed, consecutive impurities \cite{Anzurez2025}, our configuration extends this paradigm to macroscopic disorder. This enables a systematic investigation of the non-Hermitian crossover from a single localized defect to a finite concentration of randomly distributed scatterers, thereby bridging the gap between isolated impurity physics and macroscopic disorder.

The remainder of this paper is organized as follows. In Sec.~\ref{sec:model}, we introduce the detail the implementation of non-Hermitian substitutional impurities, alongside the numerical methods and spectral metrics used to characterize the system. In Sec.~\ref{sec:single_impurity}, we examine the effect of a single isolated non-reciprocal defect, demonstrating its inability to disrupt the bulk topological phase. Section~\ref{sec:symmetric} extends this analysis to a finite concentration of randomly distributed symmetric impurities, exploring the emergence of impurity mini-bands, the real-to-complex spectral transition, and the manifestation of the Non-Hermitian Skin Effect. In Sec.~\ref{sec:asymmetric}, we investigate structurally asymmetric impurity configurations, uncovering a deterministic mechanism to shift the topological phase boundaries via local defect asymmetry. Section~\ref{sec:fss} provides a finite-size scaling analysis to validate these phase boundaries, confirming a disorder-induced departure from the pristine SSH universality class. Finally, in Sec.~\ref{sec:conclusions}, we summarize our findings and outline compelling perspectives for future research.

\section{Model and methods}
\label{sec:model}
\subsection{Substitutional impurities}
We consider an ideal SSH chain consisting of $N$ unit cells, with each cell containing two distinct sublattices ($A$ and $B$). The Hamiltonian for the pristine system is given by \cite{SSH_PRL1979, Asboth_2015}:
\begin{equation}
    \hat{H} = v \sum_{m=1}^{N}\ket{m,B}\bra{m,A} + w \sum_{m=1}^{N-1} \ket{m+1,A}\bra{m,B} + h.c.
\end{equation}
where $v$ and $w$ represent the intracell (internal) and intercell (external) hopping amplitudes, respectively.

To model the presence of a localized defect, we assume a substitutional impurity at the $k$-th unit cell. This impurity modifies the local hopping landscape, leading to site-dependent amplitudes $h_i$ (intracell) and $h_e$ (intercell). To introduce non-Hermiticity, we consider non-reciprocal hopping amplitudes associated with the impurity site. Let $k$ be the index of the unit cell containing the defect. The transitions involving the impurity are governed by the following relations:
\begin{itemize}
    \item Intracell hopping ($k$-th cell): The amplitudes between sites $A$ and $B$ are defined as $h_{i, \rightarrow} = h_i + \gamma_i$ for jumps towards the impurity and $h_{i, \leftarrow} = h_i - \gamma_i$ for jumps away from it.
    \item Intercell hopping: Similarly, the connection between cell $k$ and its neighbor involves $h_{e, \rightarrow} = h_e + \gamma_e$ and $h_{e, \leftarrow} = h_e - \gamma_e$.
\end{itemize}
Here, $h_{i(e)}$ represents the symmetric hopping component, while $\gamma_{i(e)}$ quantifies the degree of non-reciprocity. This configuration effectively breaks the Hermiticity of the Hamiltonian, as the hopping matrix elements satisfy $H_{mn} \neq H_{nm}^*$. Assuming the impurity is located at site $B$ of the $k$-th cell, the local structure of the Hamiltonian in the basis $\{ \dots, \ket{k,A}, \ket{k,B}, \ket{k+1,A}, \dots \}$ is given by:
\begin{equation}
    H = \begin{pmatrix}
\ddots & v & 0 & 0 \\
v & 0 & w & 0 \\
0 & w & 0 & h_{i} + \gamma_i \\
0 & 0 & h_{i} - \gamma_i & 0 & h_{e} + \gamma_e \\
& & & h_{e} - \gamma_e & 0 & v \\
& & & 0 & v & \ddots
\end{pmatrix}. 
\end{equation}
The matrix representation highlights the broken Hermiticity, where the off-diagonal elements corresponding to the $k$-th cell are no longer complex conjugates (or transposes, in the real case). It should be noted that if the impurity were located at sublattice $A$ instead, the non-reciprocal block would appear inverted due to the bipartite structure of the SSH chain and the definition of the exterior hopping $h_e$ connecting cell $k$ with $k-1$. This structural distinction is of paramount physical importance; recent topoelectrical simulations of Hermitian SSH chains have demonstrated that the system's response, such as edge state robustness and voltage localization, is highly sensitive to the specific sublattice (A or B) occupying the impurity \cite{Anzurez2025}. Thus, preserving the precise sublattice index is crucial when extending the analysis from isolated defects to a macroscopic concentration of randomly distributed non-Hermitian scattering centers.

To contextualize our mathematical formulation, it is instructive to contrast the local Hamiltonian structure in Eq. (2) with other non-Hermitian configurations prevalent in the literature. Unlike standard models of non-Hermitian 
TAI \cite{Zhang_SciChP2020, Gong_PRX2018}, where non-reciprocity is an extensive property introduced globally across the entire chain, our model confines the asymmetric parameters ($\gamma_i, \gamma_e$) strictly to the local transitions involving the $k$-th unit cell. In this localized spirit, our approach shares fundamental similarities with recent models featuring isolated defects, such as external impurities asymmetrically coupled to a host lattice \cite{Wang_En2025, Wang_PRB2025} or non-Hermitian on-site potentials \cite{Shi_PRB2026, Hetenyi_PRB2025}, in that the defect operates as a highly sensitive, effective boundary or scattering center. However, in stark contrast to setups that couple external probes to an already non-reciprocal host \cite{Wang_En2025} or inject imaginary on-site gain/loss terms \cite{Hetenyi_PRB2025}, our substitutional approach structurally embeds the defect while preserving the perfectly Hermitian bulk and the zero on-site energies characteristic of the pristine chiral-symmetric SSH model. Ultimately, the impurity acts exclusively as a bidirectional, non-reciprocal scattering center, altering the directional flow of particles without requiring external discrete systems or a globally modified host lattice.

\subsection{Real-Space Winding Number}
The introduction of localized impurities breaks the translational symmetry of the pristine SSH chain, rendering the conventional momentum-space topological invariants ill-defined. To characterize the bulk topology of the disordered system, we employ the real-space winding number based on a spectral projection method \cite{Zhang2022,Mondragon_PRL2014,Prodan2011,Ling_PRB2021}. The real-space winding number $\omega$ is calculated as:
\begin{equation}
\omega = \frac{1}{N'} \text{Tr}' (\Gamma Q [Q, X]),
\end{equation}
where $X = \text{diag}(1, 1, 2, 2, \dots, N, N)$ represents the position operator, and $\Gamma = I_N \otimes \sigma_z$ is the chiral symmetry operator. The operator $Q$ corresponds to the homotopically equivalent flat-band Hamiltonian matrix, which is constructed from the eigenstates $|\beta\rangle$ of the system as $Q = \sum_{\beta=1}^{2N} |\beta\rangle\langle\beta| - \Gamma^{-1}|\beta\rangle\langle\beta|\Gamma$. To effectively mitigate finite-size boundary effects, the trace $\text{Tr}'$ is restricted to the middle part of the chain, tracing over the central $N' = N/2$ unit cells.

It is crucial to note that the introduction of the non-reciprocal parameters ($\gamma_i$ and $\gamma_e$) explicitly breaks the Hermiticity of the system. Consequently, the right ($|\beta_R\rangle$) and left ($\langle\beta_L|$) eigenstates of $H$ are no longer related by a simple Hermitian conjugation. To compute a topologically meaningful winding number in this non-Hermitian regime, the flat-band Hamiltonian $Q$ must be constructed using the biorthogonal basis \cite{Yao_PRL2018}:
\begin{equation}
Q = \sum_{\beta=1}^{2N} \text{sgn}(\text{Re}(E_\beta)) |\beta_R\rangle\langle\beta_L|,
\end{equation}
where $\text{Re}(E_\beta)$ ensures the states are separated by the real line gap. Numerically, this is efficiently implemented through the full eigendecomposition $H = V D V^{-1}$, where the columns of $V$ correspond to the right eigenvectors, the rows of $V^{-1}$ represent the left eigenvectors, and $D$ is the diagonal matrix containing the eigenvalues.

\subsection{Localization and Spectral Characterization}

To quantitatively characterize the spatial distribution of the eigenstates and the impact of non-Hermiticity on the energy spectrum, we introduce the Inverse Participation Ratio (IPR) and the complex energy ratio $f_\mathrm{Im}$.

The IPR for a given eigenstate $|\beta\rangle$ in the real-space basis $\{|n\rangle\}$ is defined as \cite{Zhang2022,Zhang_SciChP2020}:
\begin{equation}
    I_x^\beta = \sum_{n=1}^{2N} |\langle n|\beta\rangle|^4.
\end{equation}
This quantity serves as a measure of the degree of localization. For a state perfectly localized on a single site, $I_x^\beta = 1$, whereas for an extended state distributed uniformly across the chain of $2N$ sites, $I_x^\beta \sim 1/(2N)$. To analyze the global behavior of the system, we consider the eigenstate-averaged IPR:
\begin{equation}
    \bar{I}_x = \frac{1}{2N} \sum_{\beta=1}^{2N} I_x^\beta.
\end{equation}
In our model, $\bar{I}_x$ is essential for identifying the transition between the extended bulk phases and the localized regimes induced by a finite concentration of non-reciprocal impurities.

Furthermore, to evaluate the non-Hermitian character of the system and track the macroscopic breakdown of spectral symmetry, we define the complex energy ratio $f_\mathrm{Im}$ \cite{Zhang2021} as the fraction of eigenvalues with a non-vanishing imaginary part:
\begin{equation}
    f_\mathrm{Im} = \frac{N_{Im}}{2N},
\end{equation}
where $N_{Im}$ is the number of eigenvalues $E_\beta$ satisfying $|\text{Im}(E_\beta)| > \epsilon$, with $\epsilon=10^{-11}$ being a threshold determined by the numerical precision of the calculation. This parameter allows us to monitor the stability of the real energy spectrum and identify potential transitions to broken symmetry phases or the emergence of complex energy bands.

\subsection{Numerical Implementation}
For the numerical exact diagonalization of the system, we establish the energy scale by setting the pristine intercell hopping amplitude to $w = 1$. Furthermore, the choice of boundary conditions is of paramount importance in non-Hermitian systems due to the extreme spectral sensitivity and the potential emergence of the NHSE \cite{Yao_PRL2018, Zhang_AiP2022}. Throughout our analysis (unless explicitly mentioned), the computation of the energy spectrum, the real-space winding number $\omega$, and the espectral metrics ($\bar{I}_x$ and $f_\mathrm{Im}$) are performed under Open Boundary Conditions (OBC).

\begin{figure*}
    \centering
    \begin{subfigure}{0.32\textwidth}
        \centering
        \includegraphics[width=\linewidth]{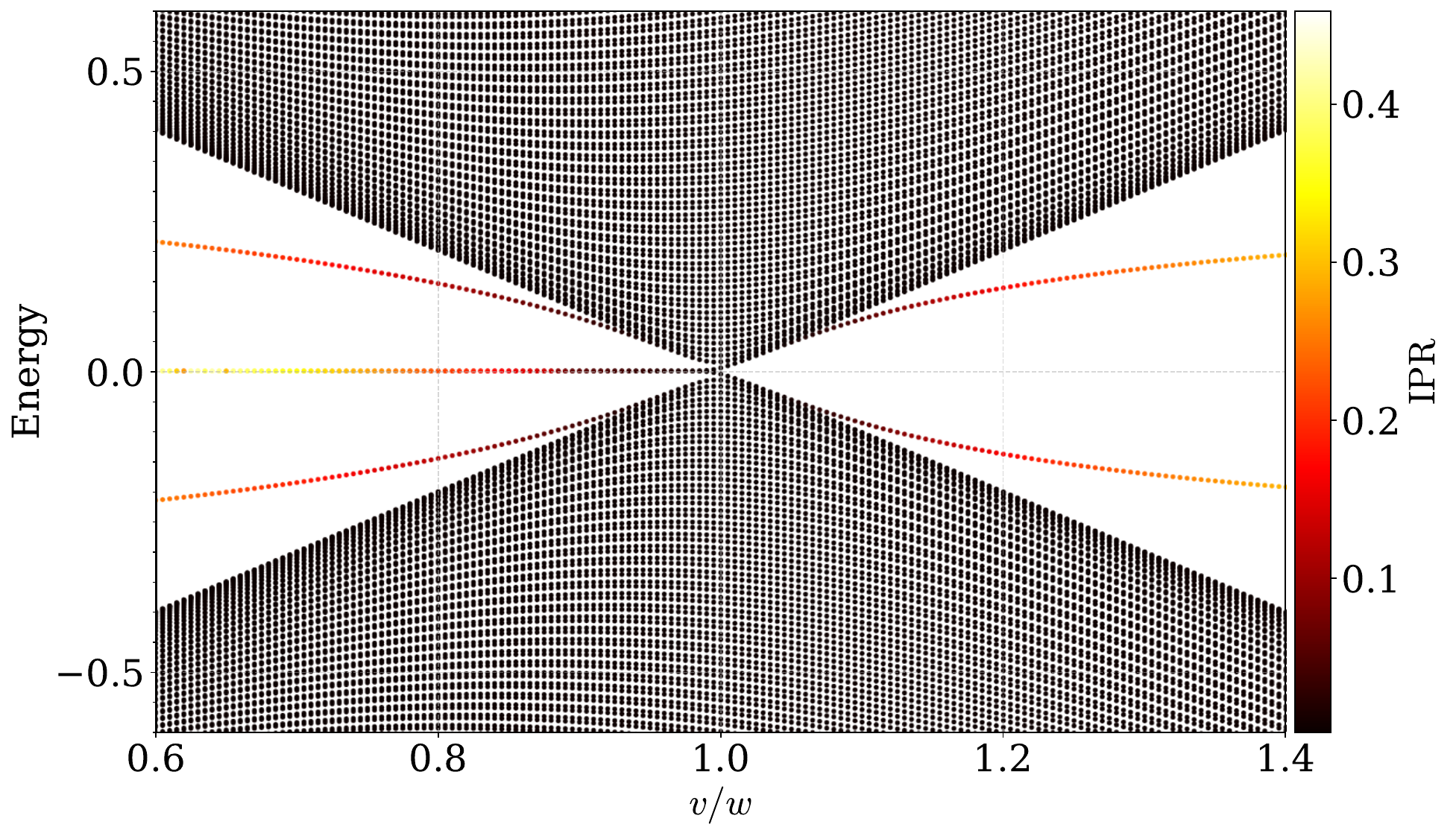}
        \caption{}
    \end{subfigure}
    \begin{subfigure}{0.32\textwidth}
        \centering
        \includegraphics[width=\linewidth]{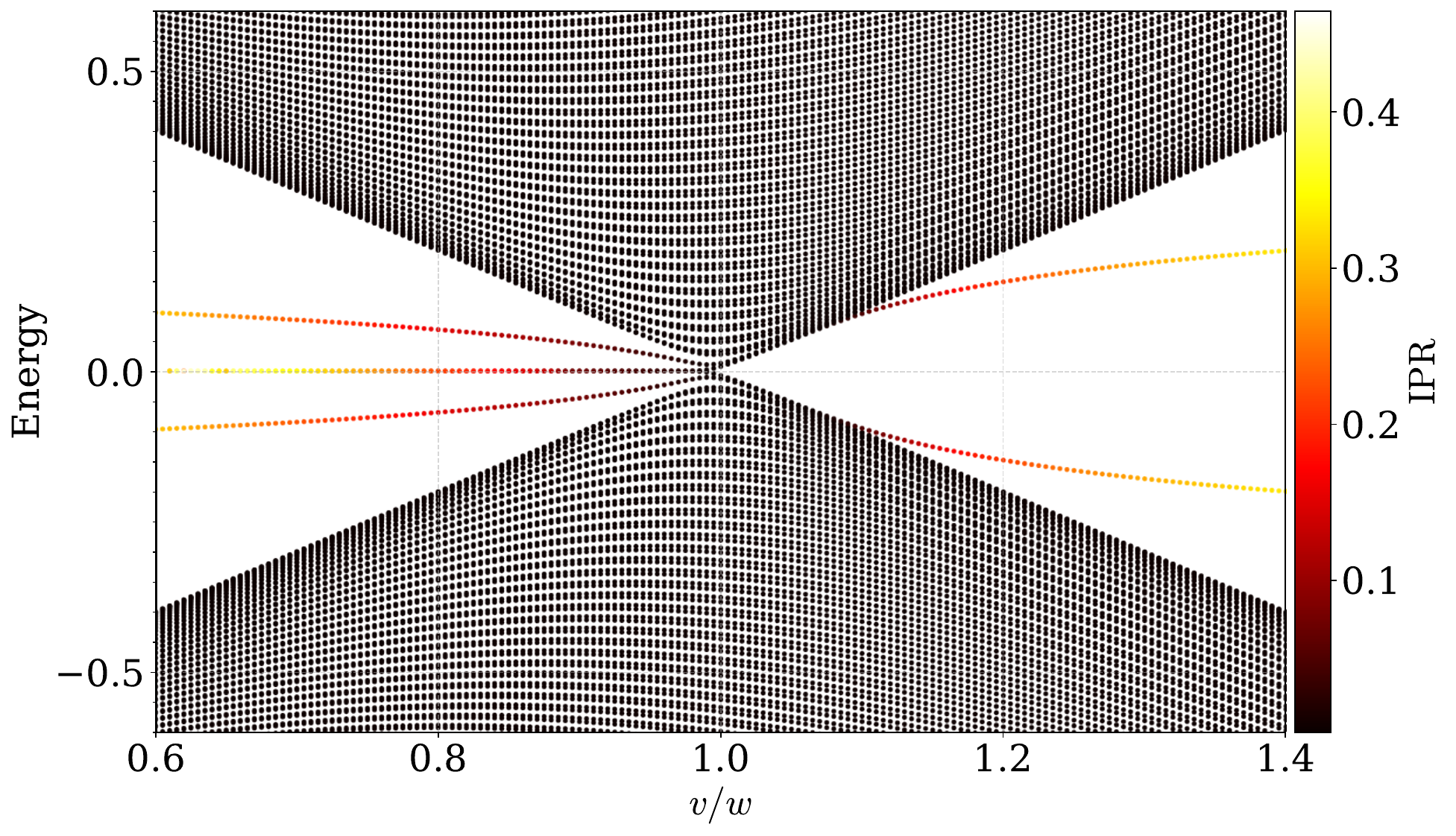}
        \caption{}
        \label{fig:spect_singleimp_m}
    \end{subfigure}    
\begin{subfigure}{0.32\textwidth}
        \centering
        \includegraphics[width=\linewidth]{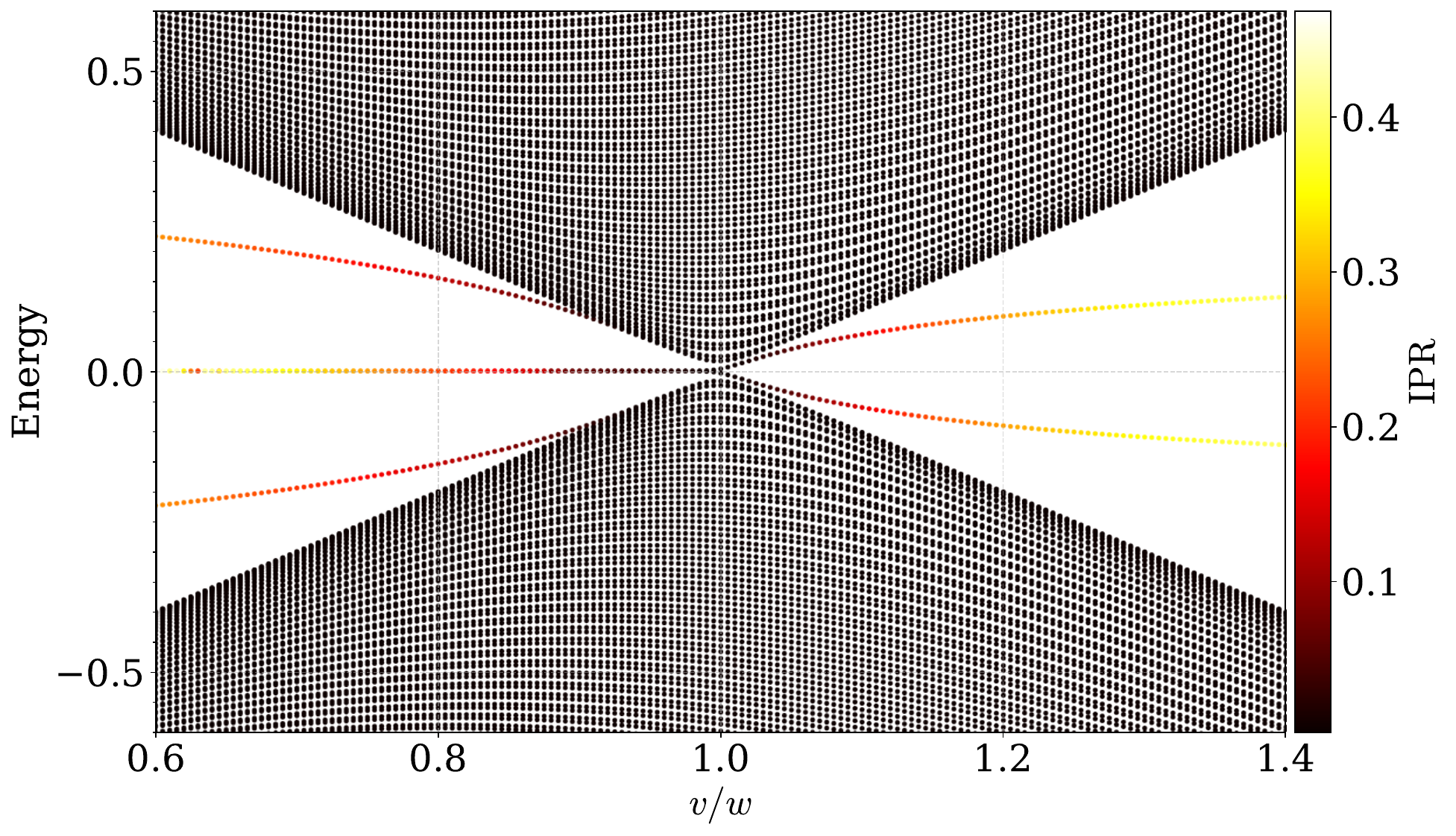}
        \caption{}
        \label{fig:spect_singleimp_r}
    \end{subfigure}

    \begin{subfigure}{0.32\textwidth}
        \centering
        \includegraphics[width=\linewidth]{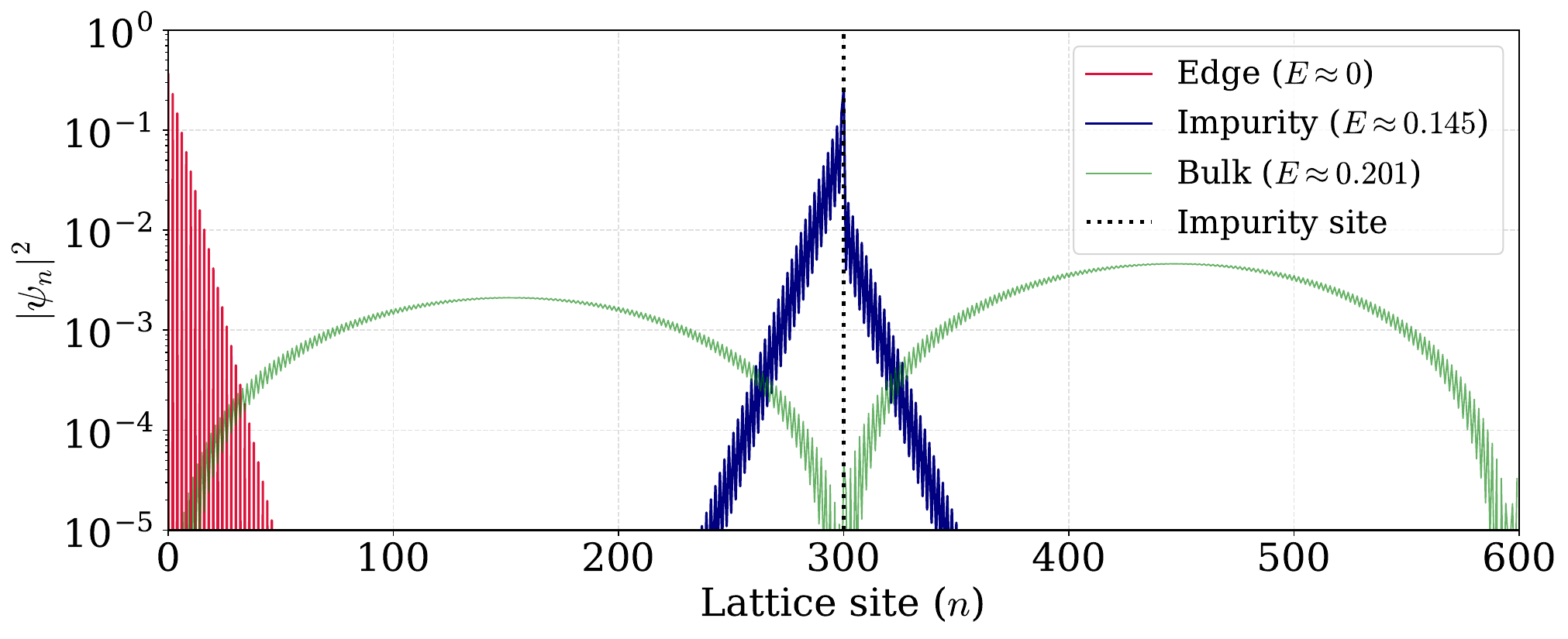}
        \caption{}
    \end{subfigure}
    \begin{subfigure}{0.32\textwidth}
        \centering
        \includegraphics[width=\linewidth]{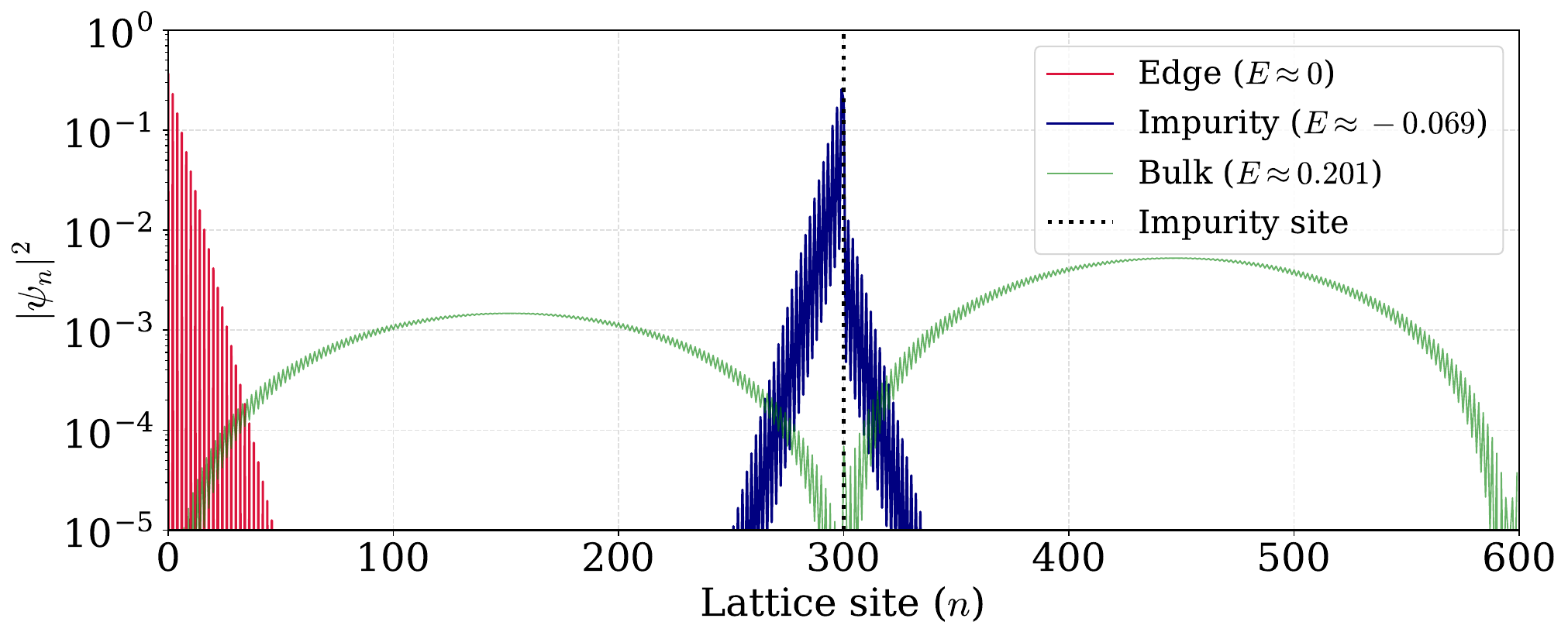}
        \caption{}
    \end{subfigure}    
\begin{subfigure}{0.32\textwidth}
        \centering
        \includegraphics[width=\linewidth]{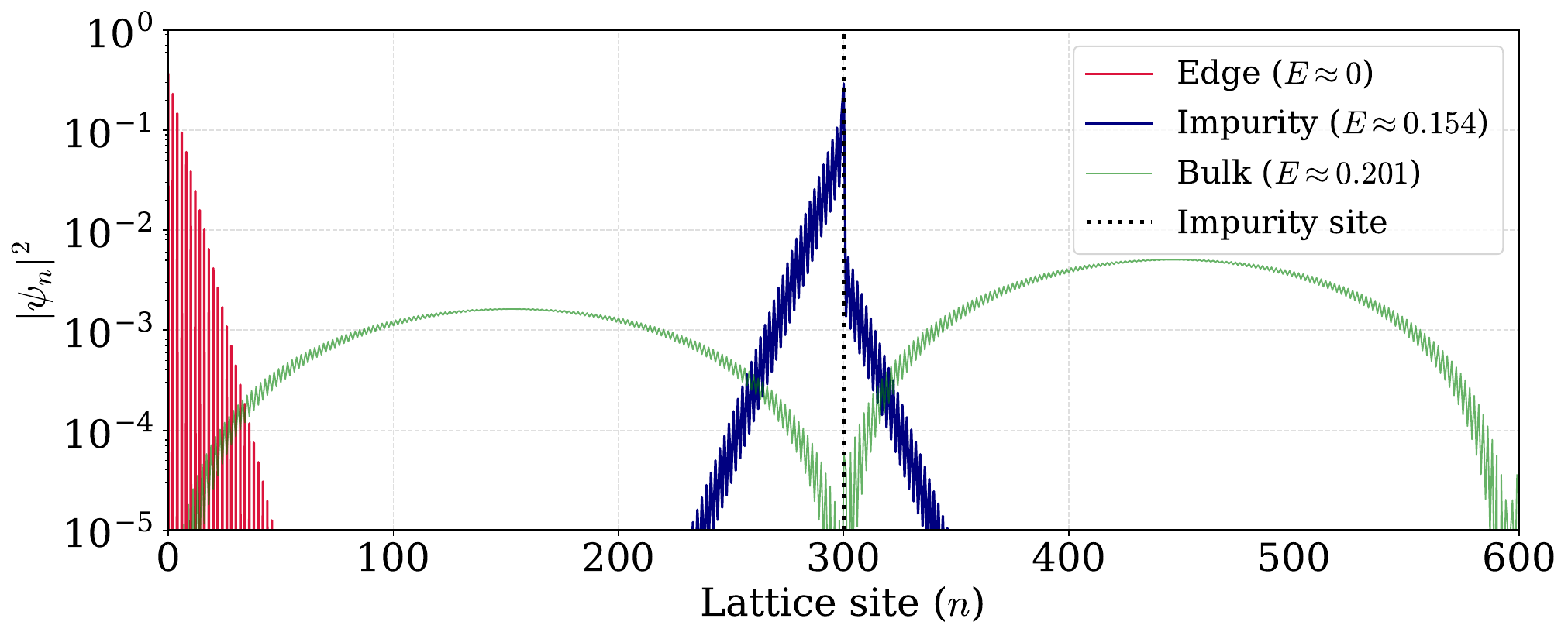}
        \caption{}
    \end{subfigure}
    \caption{Real energy spectra and spatial probability distributions for an open SSH chain ($N=300$) hosting a single non-Hermitian defect, contrasting three distinct local configurations. \textbf{(a)--(c)} Real part of the energy spectrum as a function of the hopping ratio $v/w$ ($w=1$). Eigenstates are color-coded according to their inverse participation ratio ($I_x^\beta$), resolving the degree of spatial localization across the continuum and within the gap. The top panels represent the following impurity hopping configuration: \textbf{(a)} $h_i = h_e = 0.3$ and $\gamma_i = \gamma_e = 0.05$; \textbf{(b)} $h_i = 0.15$, $h_e = 0.3$, $\gamma_i = 0.075$ and $\gamma_e = 0.05$; and \textbf{(c)} $h_i = 0.3$, $h_e = 0.2$, $\gamma_i = 0.05$ and $\gamma_e = 0.075$. \textbf{(d)--(f)} Spatial probability density profiles $|\psi_n|^2$ as a function of the lattice site $n$ evaluated at a representative hopping ratio in the topological regime ($v/w = 0.8$), displayed directly beneath their corresponding configurations in (a)--(c). Each panel contrasts a zero-energy topological edge state ($E \approx 0$), an in-gap localized impurity state ($0 < |E| < |w-v|$), and an extended bulk mode. The vertical dotted line marks the defect position at $n = N$.}
    \label{fig:spect_singleimp}
\end{figure*}

To extract the robust macroscopic properties of the system and mitigate finite-size artifacts inherent to specific random defect configurations, we systematically compute disorder-averaged observables. For the analysis of macroscopic disorder, we set the system size to $N = 300$ unit cells, maintaining an optimal balance between numerical resolution and computational efficiency. Physical quantities shown in two-dimensional phase diagrams are averaged over an ensemble of $N_r=250$ independent random realizations of the impurity distribution for each parameter set. For one-dimensional profiles and scaling analyses requiring higher statistical clarity, the ensemble size is further increased to ensure rigorous statistical convergence. This extensive ensemble averaging effectively suppresses configuration-specific noise and reveals the typical behavior of the topological phase transitions and localization properties, consistent with established methodologies in the study of disordered topological insulators \cite{Zhang2022, Titum2015, Prodan2011}.

\section{Single impurity case}
\label{sec:single_impurity}

We begin by examining the effect of a single substitutional impurity on the SSH chain. Figure~\ref{fig:spect_singleimp} illustrates the OBC energy spectrum (top row) and spatial probability densities (bottom row) for a finite chain ($N=300$), contrasting symmetric and asymmetric hopping configurations. 

The shaded spectral regions corresponding to extended bulk states (low IPR) confirm that a single localized defect cannot macroscopically alter the extended nature of the bulk modes. Instead, the non-Hermitian perturbation induces highly localized, impurity-bound midgap states. Structural asymmetry systematically shifts the energy of these states exclusively within either the topological ($h_i$ modifies energies for $v<w$ in central panel) or trivial ($h_e$ modifies energies for $v>w$ in right panel) regimes. In general, increasing the symmetric hopping h raises the energy of the impurity states, driving them back into the bulk if the local coupling is sufficiently strong. In the extreme non-Hermitian limit ($h=\gamma$), the backward transition is entirely suppressed, effectively severing the chain and inducing a strict boundary mode. Crucially, as the system approaches the topological phase transition ($v \approx w$), these defect states merge into the gapless continuum, demonstrating that an isolated impurity cannot destroy the macroscopic phase boundary. 

The $|\psi_n|^2$ profiles clearly show the impurity acting as a strong scattering center, confining the midgap state around $n = N$ and inducing a sharp spatial discontinuity, while preserving the extended, wave-like envelope of the bulk state. The varying degrees of discontinuity across the three configurations indicates that tuning the local non-reciprocal parameters effectively modulates the scattering cross-section of the defect. Throughout all single-impurity configurations, the energy spectrum remains entirely real, and the real-space winding number remains strictly quantized and identical to the pristine case.

This resilience contrasts sharply with recent findings in other tight-binding models, where a single non-Hermitian impurity can act as a macroscopic sink, triggering a local NHSE and forcing the emergence of complex eigenenergies \cite{Hetenyi_PRB2025, Spring_SciPost2024}, or sequentially decompose the global topological invariant \cite{Shi_PRB2026}. Moreover, in simple Hermitian lattices, a single on-site potential is sufficient to drive a localization transition \cite{Khan_JLTP2026}. In our chiral-symmetric SSH model, the robust topological gap effectively quarantines the non-Hermiticity, confining its effects strictly to the in-gap defect states and preserving the extended nature of the bulk modes. This inherent topological suppression shields the global winding number from isolated non-reciprocity and prevents macroscopic spectral collapse. Therefore, the global breakdown of topology and the onset of complex energies in this system strictly require a finite, macroscopic concentration of non-reciprocal scatterers.

\section{Finite Concentration of Random Impurities: Symmetric configuration}
\label{sec:symmetric}

Having established the effect of a single localized non-Hermitian defect on the SSH chain, we now consider a finite concentration of such impurities, randomly distributed throughout the bulk of the lattice. The impurity density $r$ is  defined as the ratio of the number of impurities $n_i$ to the number of lattice sites $2N$, such that $r = n_i / (2N)$. This random distribution introduces multiple scattering centers that explicitly break both translational symmetry and Hermiticity across the entire system. We systematically compute the spectral and topological properties of the system as a function of the pristine hopping ratio $v/w$ and the impurity concentration $r$ to evaluate the robustness of the system against macroscopic non-Hermitian disorder.

\begin{figure}[htpb]
    \centering
    \begin{subfigure}{0.8\linewidth}
        \centering
        \includegraphics[width=\linewidth]{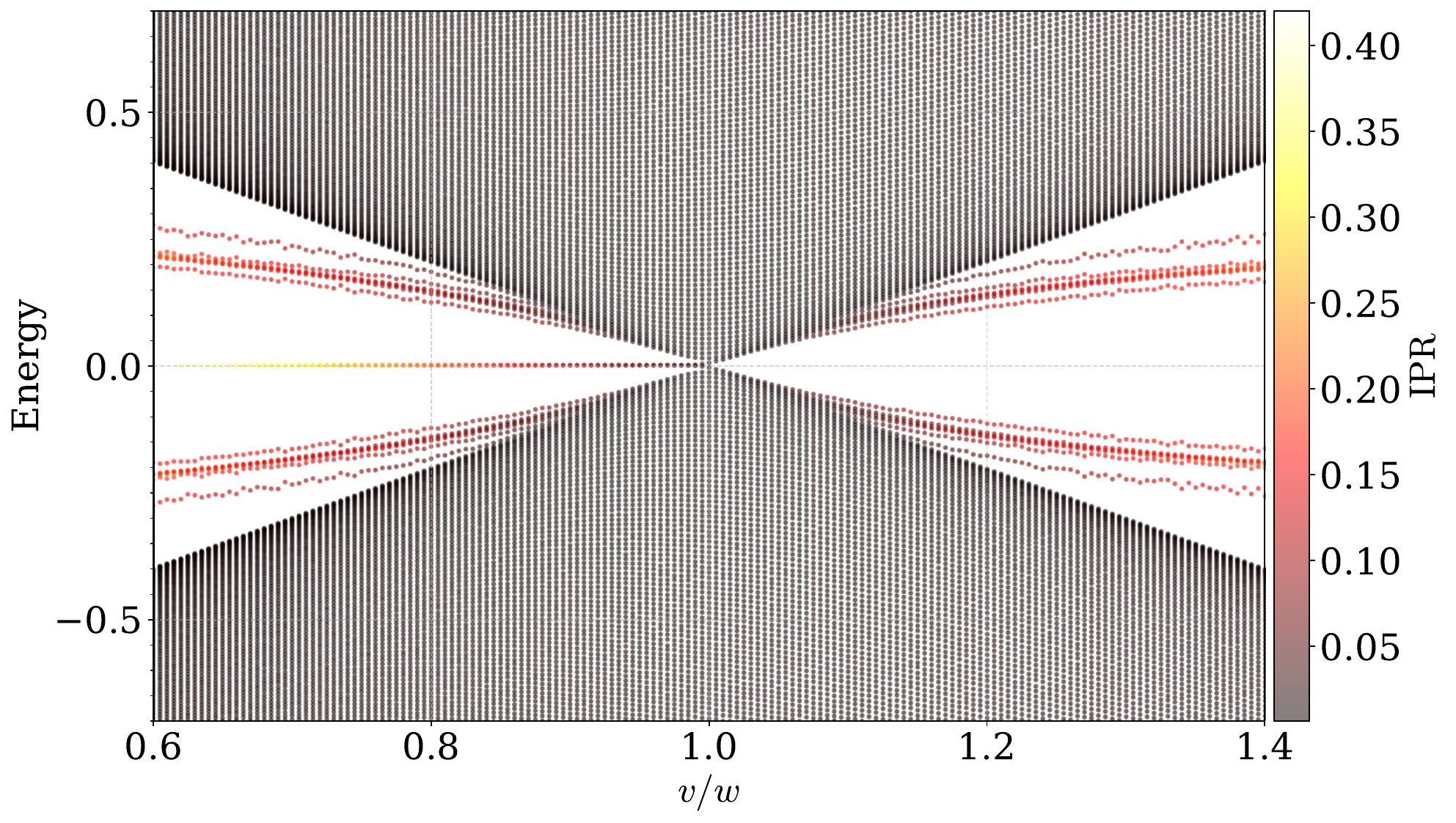}\vspace{-.23cm}
        \caption{}
    \end{subfigure}\vspace{-.18cm}
    
    \begin{subfigure}{0.8\linewidth}
        \centering
        \includegraphics[width=\linewidth]{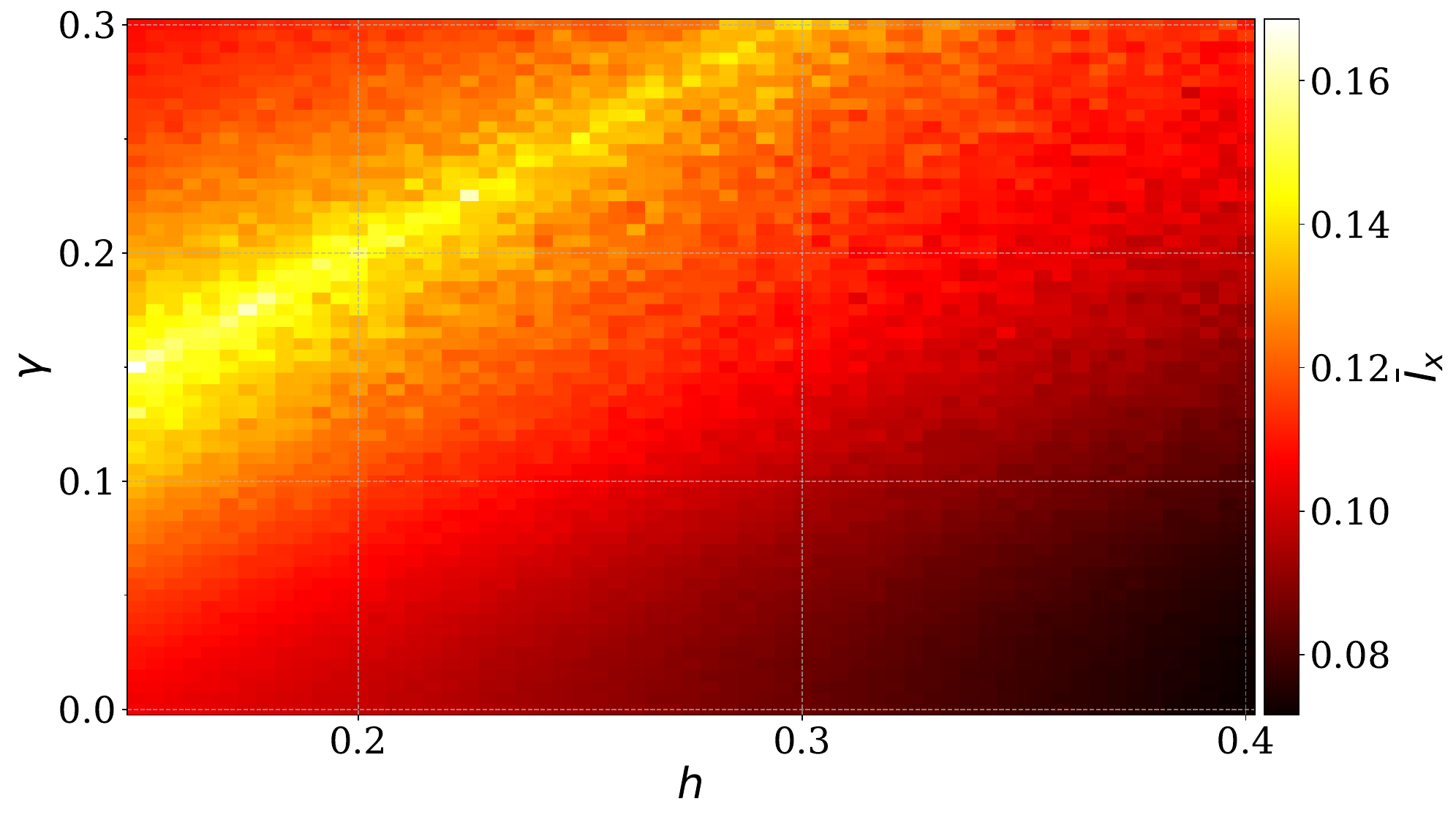}\vspace{-.23cm}
        \caption{}
    \end{subfigure}\vspace{-.18cm}
    
    \begin{subfigure}{0.8\linewidth}
        \centering        \includegraphics[width=\linewidth]{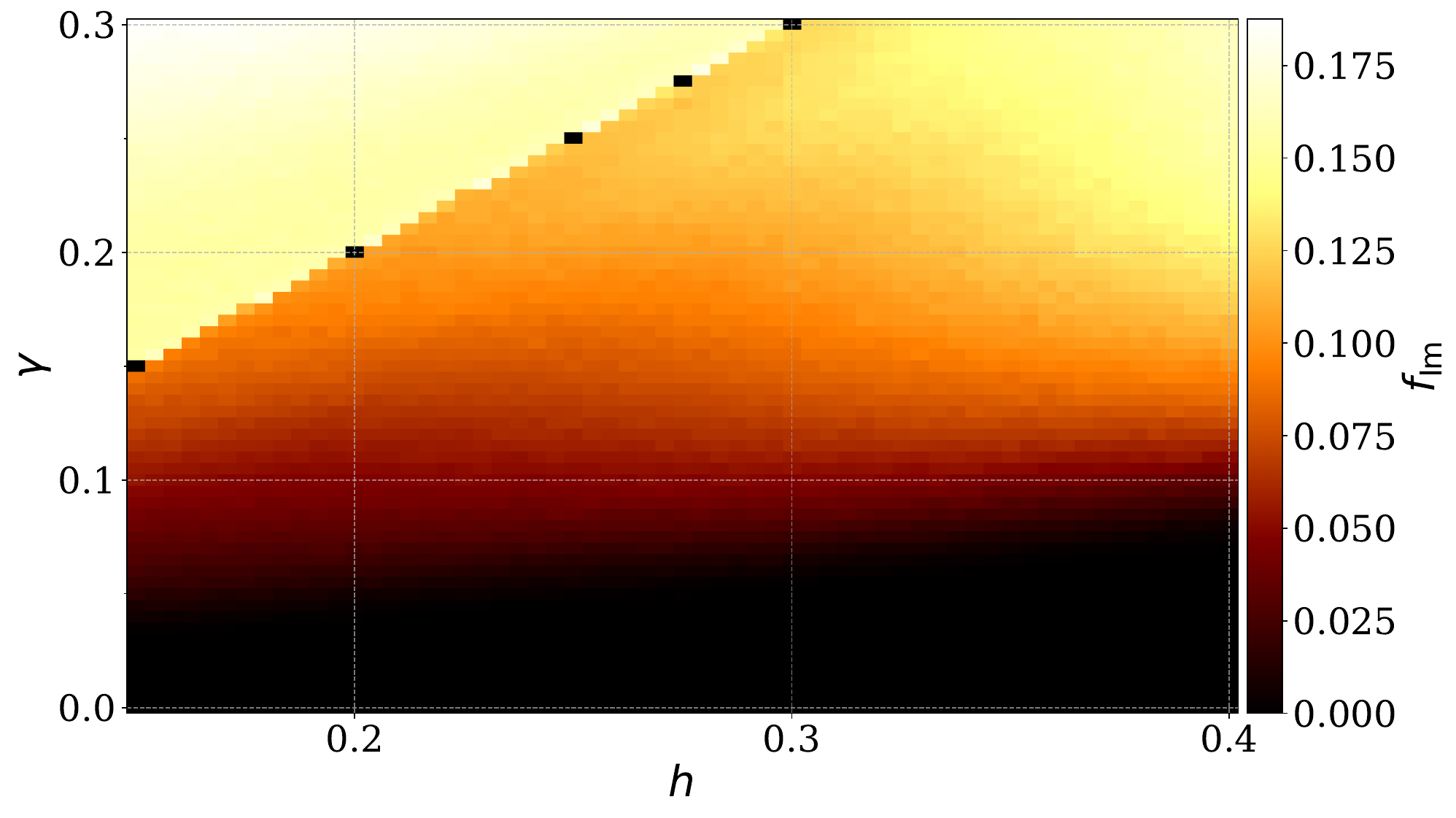}\vspace{-.23cm}
        \caption{}
    \end{subfigure}\vspace{-.18cm}
    
    \caption{Spectral and localization properties of the SSH chain subjected to a finite concentration of symmetrically configured non-Hermitian impurities ($h_i = h_e \equiv h$, $\gamma_i = \gamma_e \equiv \gamma$). \textbf{(a)} Disorder-averaged real part of the energy spectrum ($N_r = 150$) as a function of $v/w$, evaluated at $r = 1\%$ with $h = 0.3$ and $\gamma = 0.05$. Eigenstates are color-coded by the corresponding $I_x^\beta$, revealing the formation of highly localized impurity mini-bands within the energy gap. \textbf{(b)} and \textbf{(c)} display heat maps of the disorder-averaged $\bar{I}_x$ and $f_{\mathrm{Im}}$, respectively, mapped across the $(h, \gamma)$ parameter space at $r = 10\%$ and $v/w=1$ ($N_r = 250$).}

    \label{fig:macro_symmetric}
\end{figure}

We begin by examining the structurally symmetric case, where both intracell and intercell impurity transitions are identical ($h_i = h_e \equiv h$ and $\gamma_i = \gamma_e \equiv \gamma$). Figure~\ref{fig:macro_symmetric}(a) shows the disorder-averaged energy spectrum in the dilute impurity limit ($r = 1\%$), with eigenstates color-coded according to $I_x^\beta$. As the spatial distance between defects decreases, their exponentially localized wavefunctions overlap and hybridize. This interaction broadens the isolated defect modes into highly localized impurity mini-bands within the bulk energy gap. Crucially, these mini-bands are formed by displacing states from the bulk; as impurities are introduced, extended modes are systematically drawn into the localized in-gap manifold. This proliferation of highly confined states (yellow/red markers) provides a macroscopic physical realization of the impurity-induced topological decomposition recently proposed by Shi \textit{et al.} \cite{Shi_PRB2026}. The non-Hermitian defects effectively fragments the chain into smaller topological segments, sequentially degrading the global winding topology as the impurity bands accumulate. While these localized bands will eventually proliferate and cluster near the critical point at higher densities, in this dilute regime ($r = 1\%$) with moderate non-reciprocity ($\gamma = 0.05$), the energy spectrum remains predominantly real, preserving the overall spectral stability of the system.

To map the boundaries of this spectral stability, we increase the defect concentration to a macroscopic density of $r = 10\%$ and explore the broader $(h, \gamma)$ parameter space. Figures~\ref{fig:macro_symmetric}(b) and \ref{fig:macro_symmetric}(c) display heat maps for the disorder-averaged $\bar{I}_x$ and $f_{\mathrm{Im}}$, respectively. As seen in panel (b), $\bar{I}_x$ exhibits a prominent, bright diagonal ridge of maximal localization, which traces the $h = \gamma$ diagonal. Correspondingly, in panel (c), while $f_{\mathrm{Im}}$ exhibits a gradual increase prior to this threshold, the $h = \gamma$ diagonal perfectly coincides with an abrupt jump where the complex spectrum fraction saturates and stabilizes at uniformly high values. Physically, this precise correlation at $h = \gamma$ reveals the underlying mechanism of the spectral complex saturation: the transition probability in one direction vanishes entirely ($h - \gamma = 0$), effectively dimerizing the chain, with the defects acting as internal effective boundaries \cite{Wang_En2025, Shi_PRB2026}, naturally increasing spatial confinement.

\begin{figure*}
    \centering
    \begin{subfigure}{0.32\textwidth}
        \centering
        \includegraphics[width=\linewidth]{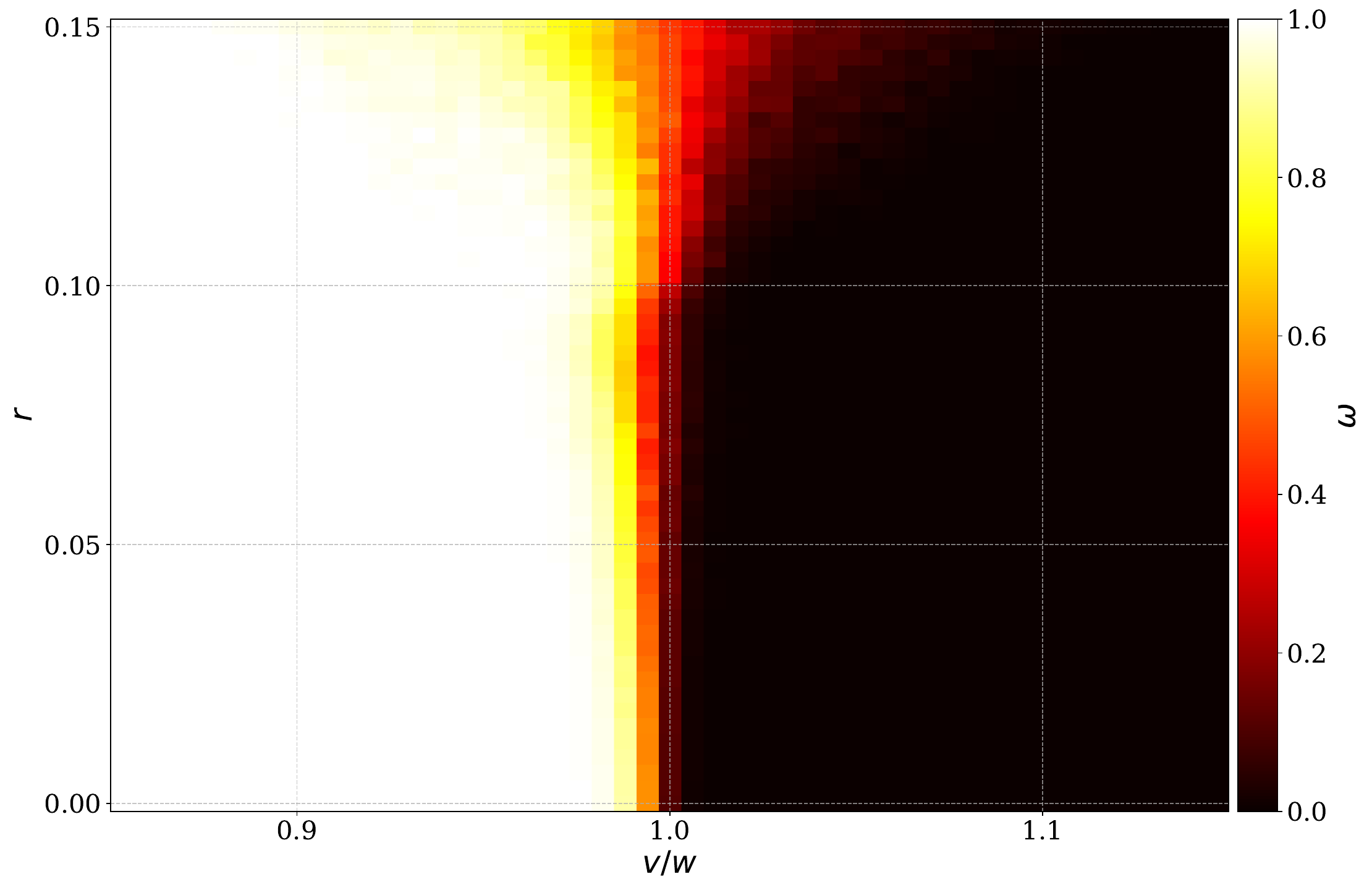}
        \caption{}
    \end{subfigure}
    \begin{subfigure}{0.32\textwidth}
        \centering
        \includegraphics[width=\linewidth]{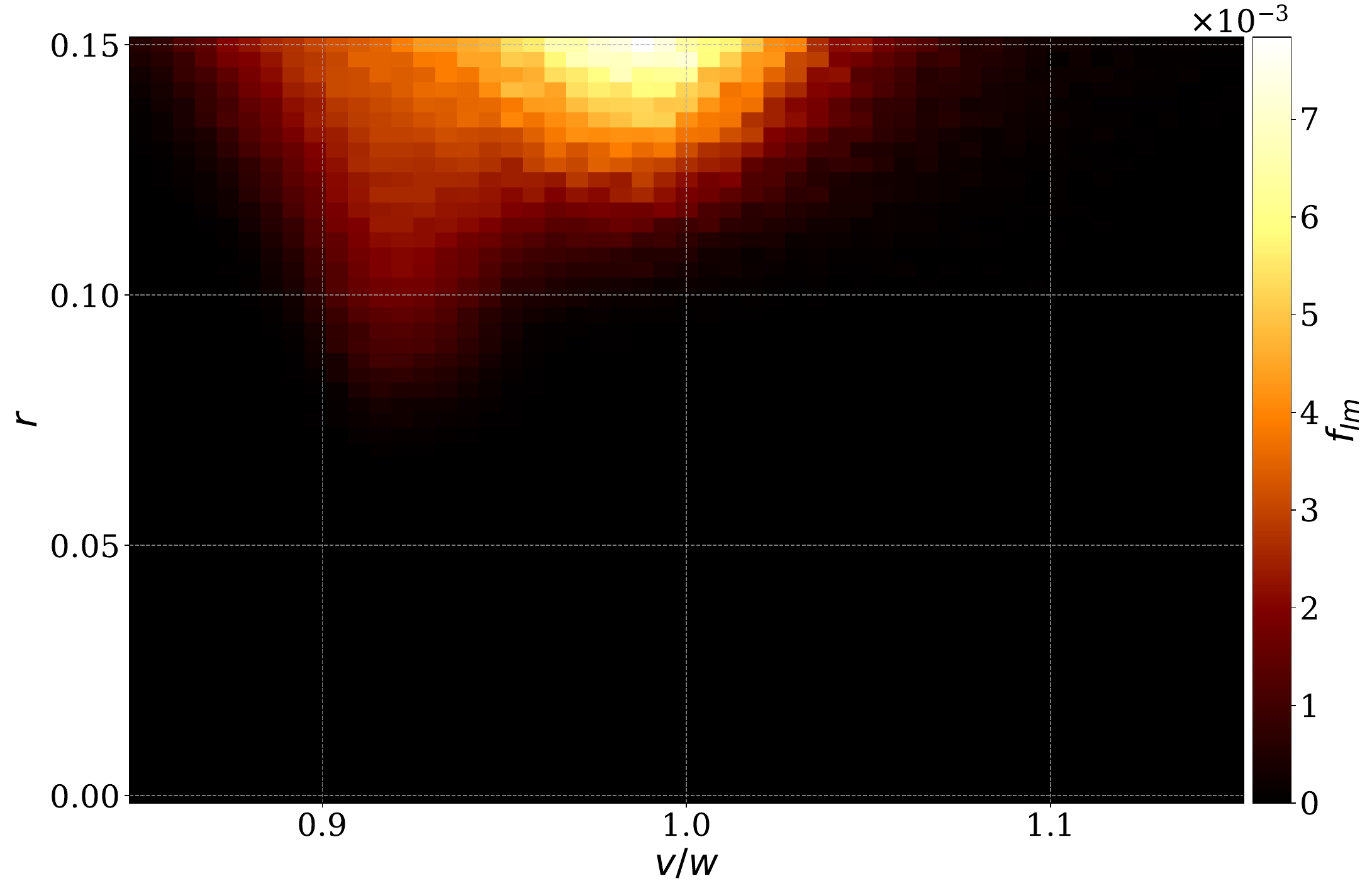}
        \caption{}
    \end{subfigure} 
    \begin{subfigure}{0.32\textwidth}
        \centering
        \includegraphics[width=\linewidth]{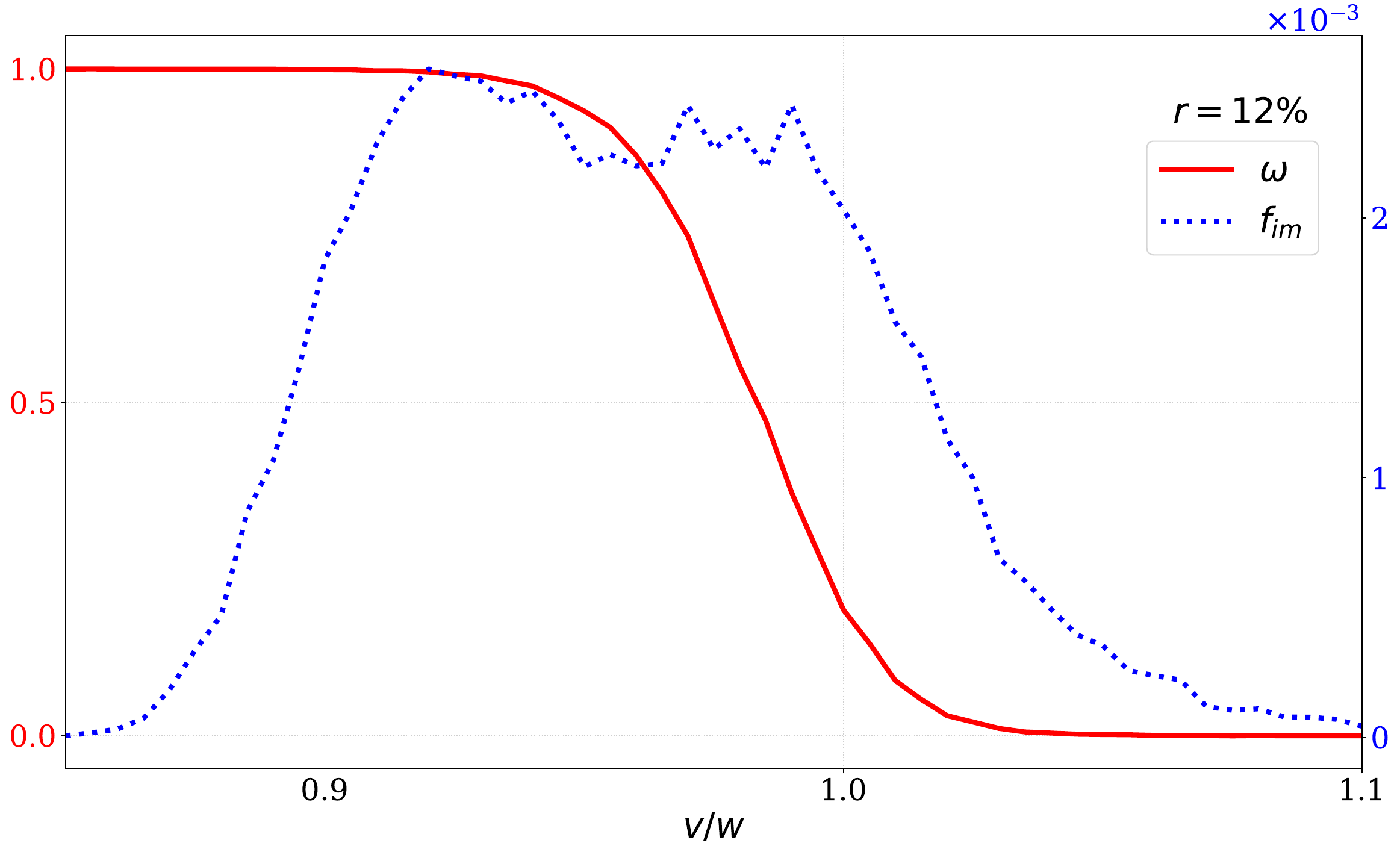}

        \caption{}
    \end{subfigure}

    \begin{subfigure}{0.32\textwidth}
        \centering
        \includegraphics[width=\linewidth]{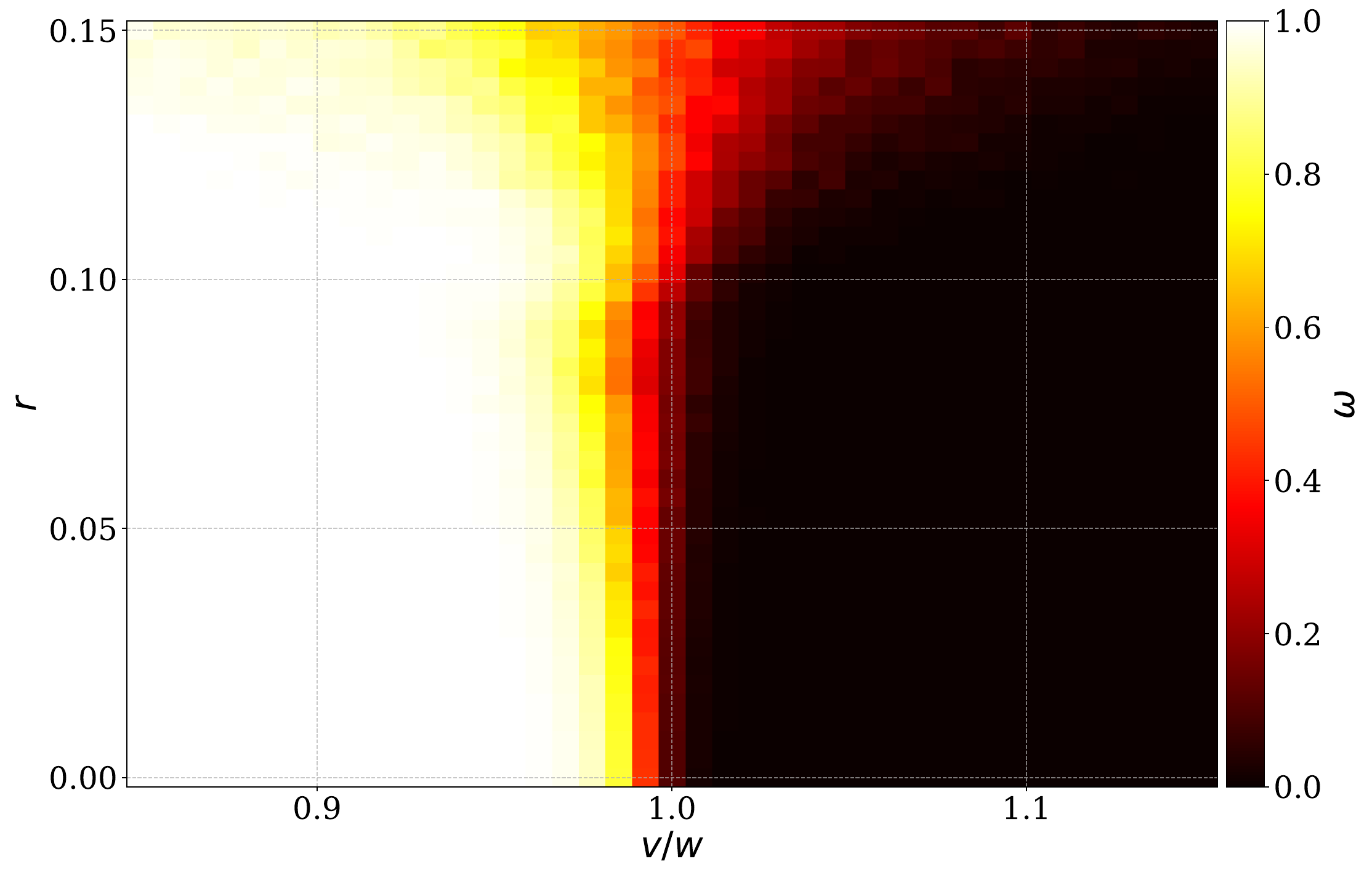}
        \caption{}
    \end{subfigure}
    \begin{subfigure}{0.32\textwidth}
        \centering
        \includegraphics[width=\linewidth]{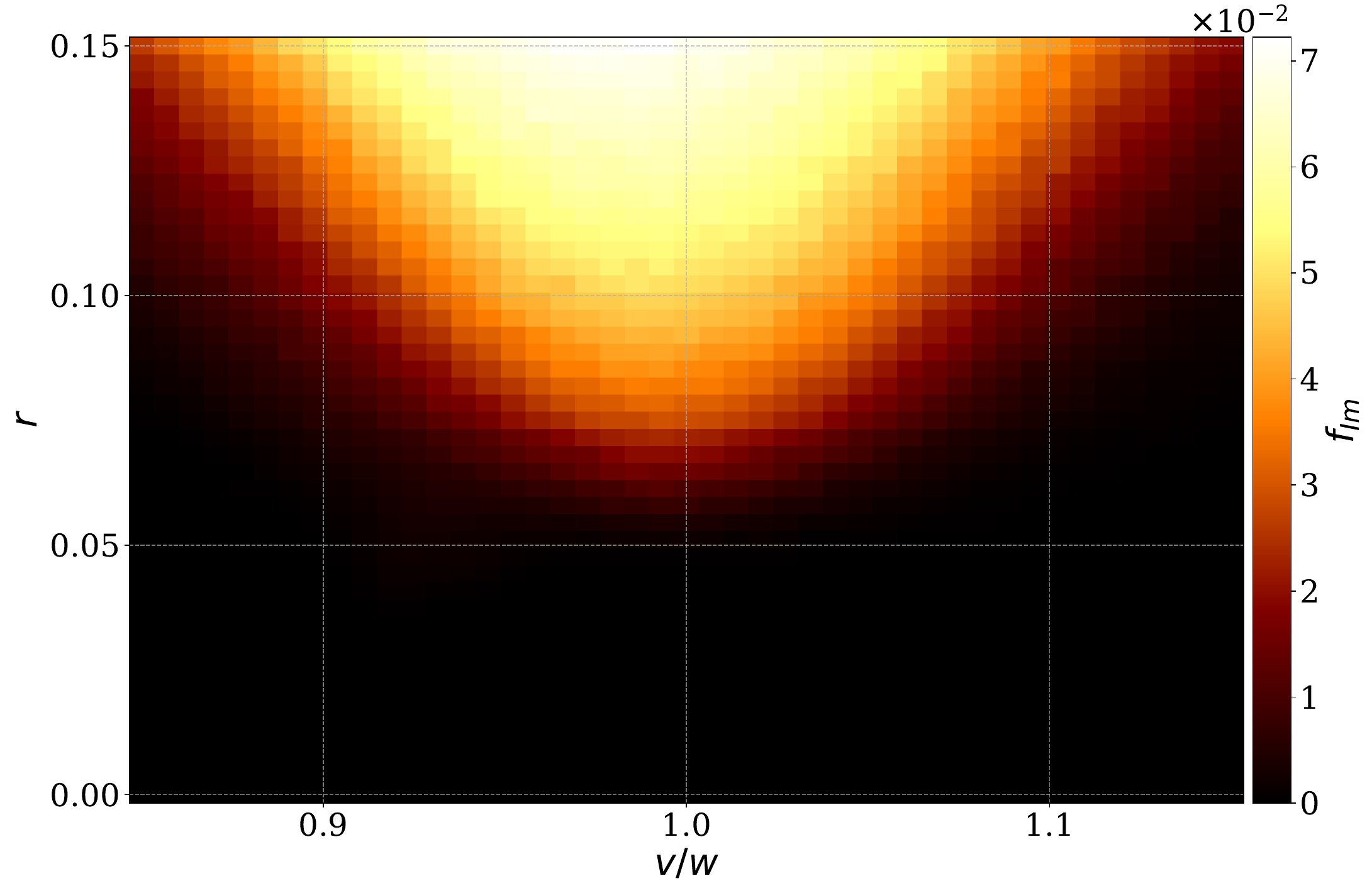}
        \caption{}
    \end{subfigure}    
\begin{subfigure}{0.32\textwidth}
        \centering
        \includegraphics[width=\linewidth]{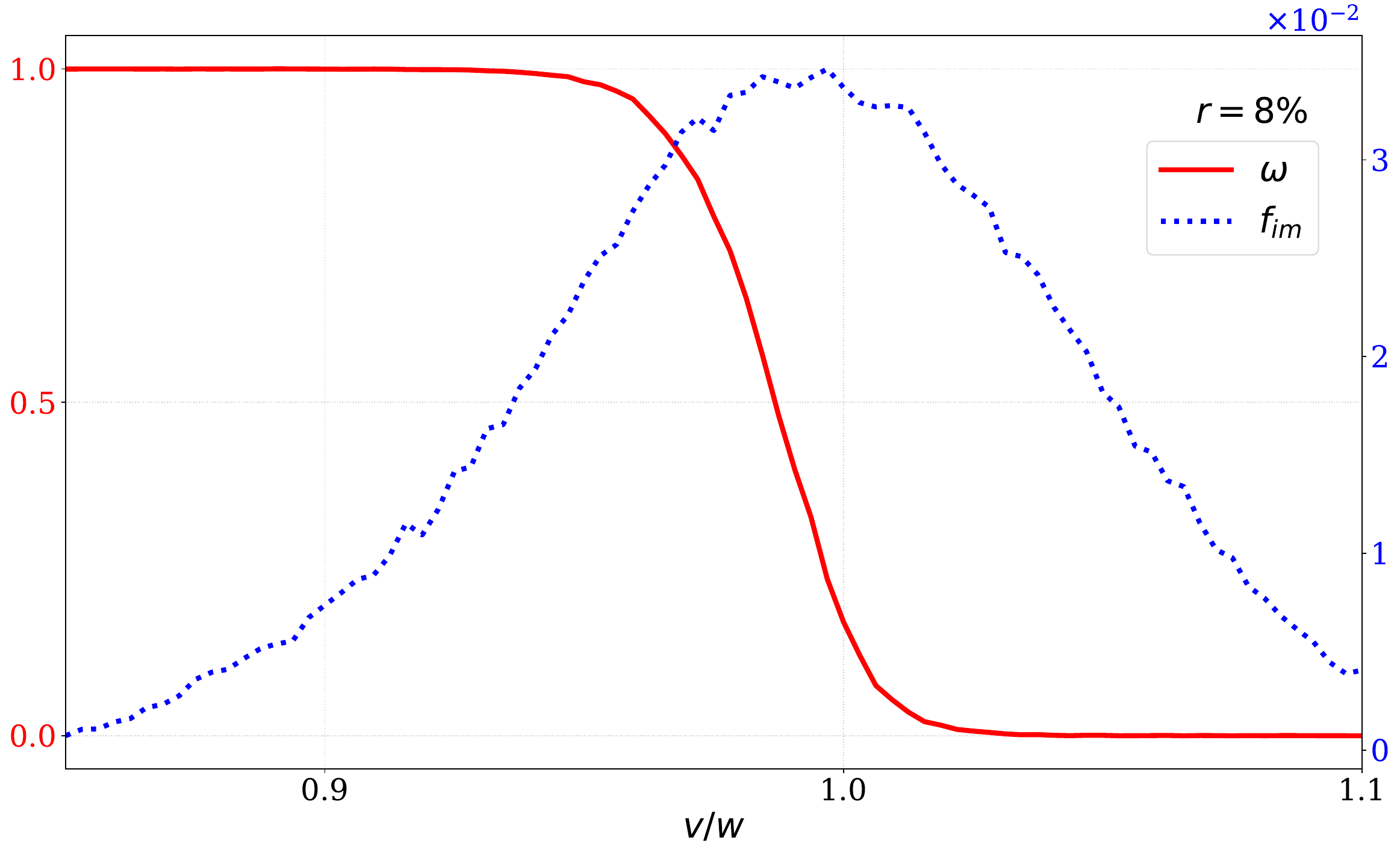}
        \caption{}
        \label{fig:average_2D_M_p2}
    \end{subfigure}
    \caption{Interplay between topology and non-Hermiticity in the disordered SSH chain. The structurally symmetric impurity parameters are fixed at $h = 0.3$, while the non-reciprocal strengths are set to $\gamma = 0.05$ (top row) and $\gamma = 0.10$ (bottom row). The left and middle columns display heat maps of the disorder-averaged real-space winding number $\omega$ ($N_r = 250$) and the complex spectrum fraction $f_{\mathrm{Im}}$ ($N_r = 750$), respectively, mapped across the $(v/w, r)$ parameter space. The rightmost column presents horizontal cross-sections of these phase diagrams at $r=12\%$ and $r=8\%$ respectively ($N_r = 1000$). The numerical noise visible in \textbf{c)} reflects the difficulty of achieving statistical convergence due to extremely sparse complex eigenenergies.}
    \label{fig:phase_diagrams_sym}
\end{figure*}

However, as non-reciprocity is increased further ($\gamma > h$), the backward transition amplitude ($h - \gamma$) becomes negative and $\bar{I}_x$ systematically decreases. As recently demonstrated for isolated non-Hermitian defects \cite{Hetenyi_PRB2025}, such a sign reversal between directional hoppings fundamentally alters the local scattering dynamics and directly forces the emergence of complex eigenenergies. This mechanism drives the system deep into a heavily complex spectral phase where $f_{\mathrm{Im}}$ saturates. Because this regime introduces distinct physics associated with these negative transition amplitudes, we will strictly confine our subsequent analysis of the disorder-driven topological phase transitions to the $\gamma < h$ parameter space, where the gradual smearing of the topological phase and the stability of the real energy gap can be systematically evaluated.

We now investigate the topological phase diagrams across the $(v/w, r)$ parameter space to evaluate the disorder-driven smearing of the phase boundary. We fix the symmetric local hopping at $h = 0.3$ and contrast two distinct non-reciprocal strengths: $\gamma = 0.05$ (top row) and $\gamma = 0.10$ (bottom row) in Fig.~\ref{fig:phase_diagrams_sym}. The left column displays the disorder-averaged real-space winding number $\omega$. In the clean limit ($r \to 0$), the topological transition occurs near $v/w = 1$, with a small finite width strictly attributable to finite-size effects. However, as $r$ increases, this transition region visibly broadens, capturing the macroscopic smearing of the topological phase boundary caused by the hybridization of the dense impurity mini-bands with the bulk continuum. Crucially, comparing the top and bottom rows reveals that elevating $\gamma$ significantly lowers the threshold density $r$ required to trigger this broadening, thus acting as a catalyst, weakening the topological gap and rendering the system more susceptible to macroscopic disorder.

This erosion of the mobility gap is directly related to the breakdown of Hermiticity, as observed in the $f_{\mathrm{Im}}$ heat map (middle column). Complex eigenenergies only emerge at high r, mapping out a growing region that mirrors that of the smeared transition of $\omega$. For $\gamma = 0.05$, the onset of complex energies requires $r > 10\%$, and its maximum value remains on the order of $\sim 7 \times 10^{-3}$. Given the lattice size of $N=300$, this indicates that fewer than a dozen states acquire imaginary components. Because this subset of complex eigenvalues is exceedingly sparse, achieving smooth statistical convergence for the ensemble average is highly computationally demanding. Consequently, even with an enhanced ensemble size of $N_r = 1000$ used for the one-dimensional cross-sections, residual numerical noise remains visible in the $f_{\mathrm{Im}}$ profile shown in Fig.~\ref{fig:phase_diagrams_sym}c). In stark contrast, for $\gamma = 0.10$, the real-to-complex transition is triggered just above $r = 5\%$, and the maximum $f_{\mathrm{Im}}$ increases by a full order of magnitude ($\sim 7 \times 10^{-2}$). This confirms that while the topological gap protects the real spectrum at low disorder, enhanced non-reciprocal pumping drastically accelerates the macroscopic breakdown of spectral Hermiticity once the gap begins to close.

The synchronization between topological and spectral boundaries is highlighted in the one-dimensional horizontal cross-sections at $r = 12\%$ (right column). These profiles demonstrate that the continuous degradation of the winding number coincides with the surge of the complex spectrum fraction. Importantly, this behavior provides a clear contrast to the TAI phases frequently reported in both Hermitian \cite{Li_PRL2009} and non-Hermitian \cite{Zhang_SciChP2020} frameworks. While global non-Hermiticity and uncorrelated on-site disorder can synergistically generate robust non-trivial topology from trivial insulators \cite{Zhang_SciChP2020}, the symmetric macroscopic disorder in our model acts predominantly as a destructive perturbation that systematically erodes chiral symmetry. 

\begin{figure}[htpb]
    \centering
    \begin{subfigure}{0.245\textwidth}
        \includegraphics[width=\linewidth]{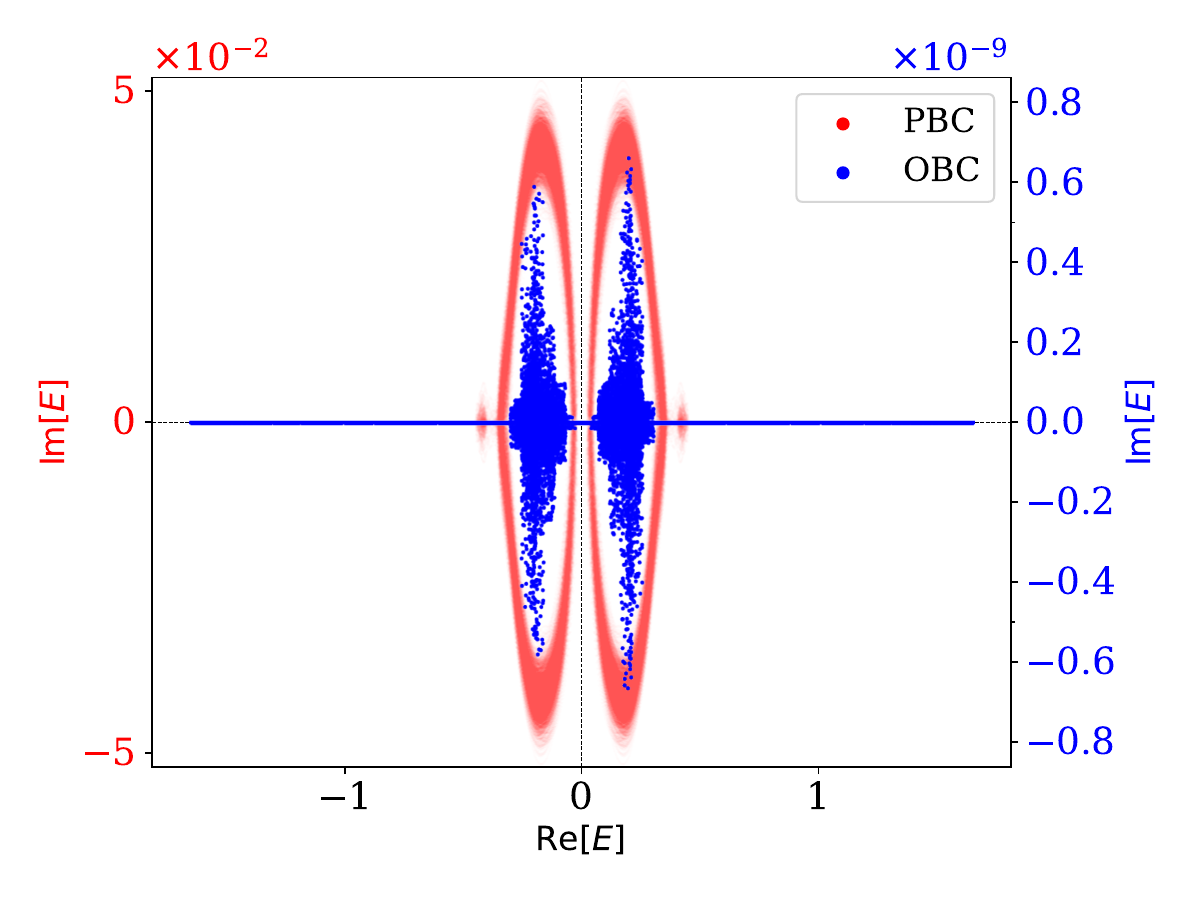}
        \caption{}
    \label{fig:spectra_symv065} \end{subfigure}\hspace{-.22cm}
    \begin{subfigure}{0.245\textwidth}
        \includegraphics[width=\linewidth]{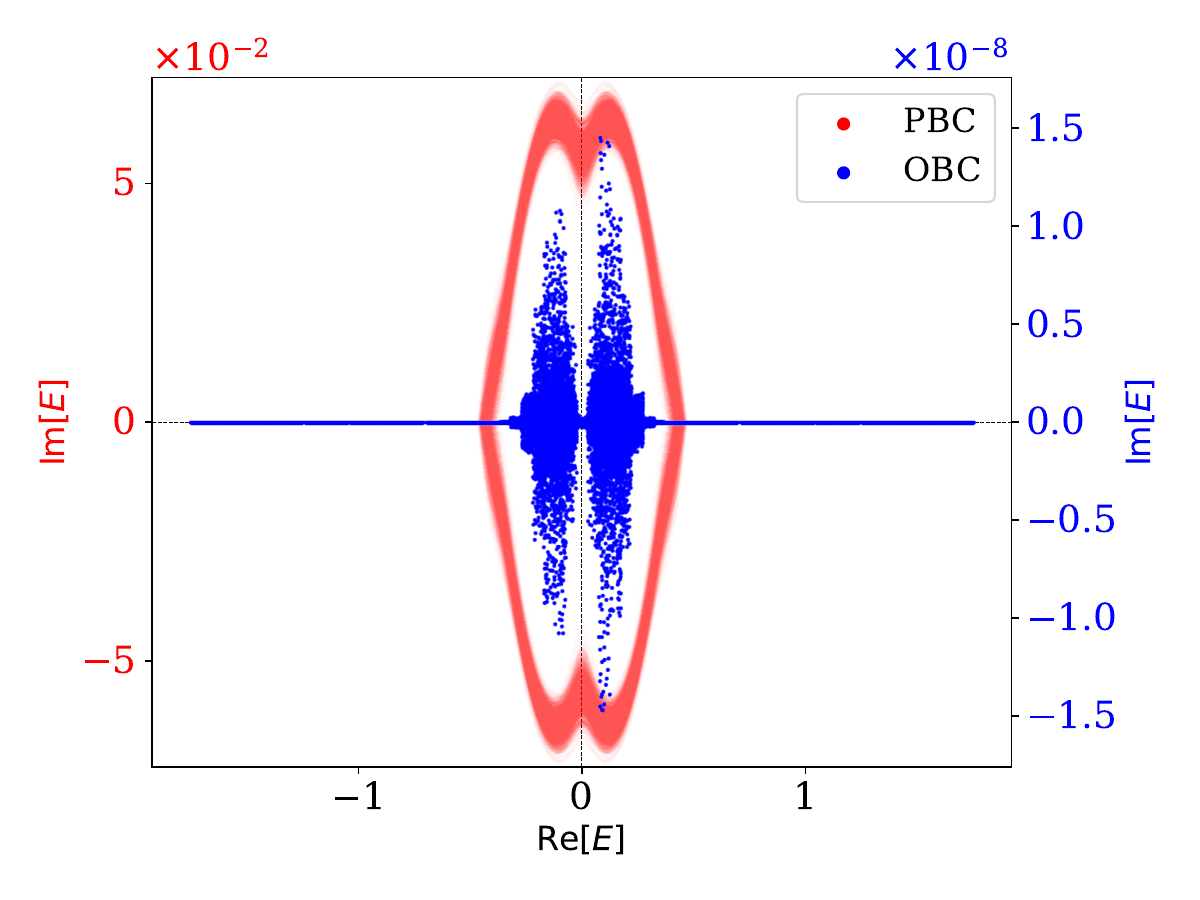}
    \caption{}\end{subfigure}
    
    \begin{subfigure}{0.245\textwidth}
        \includegraphics[width=\linewidth]{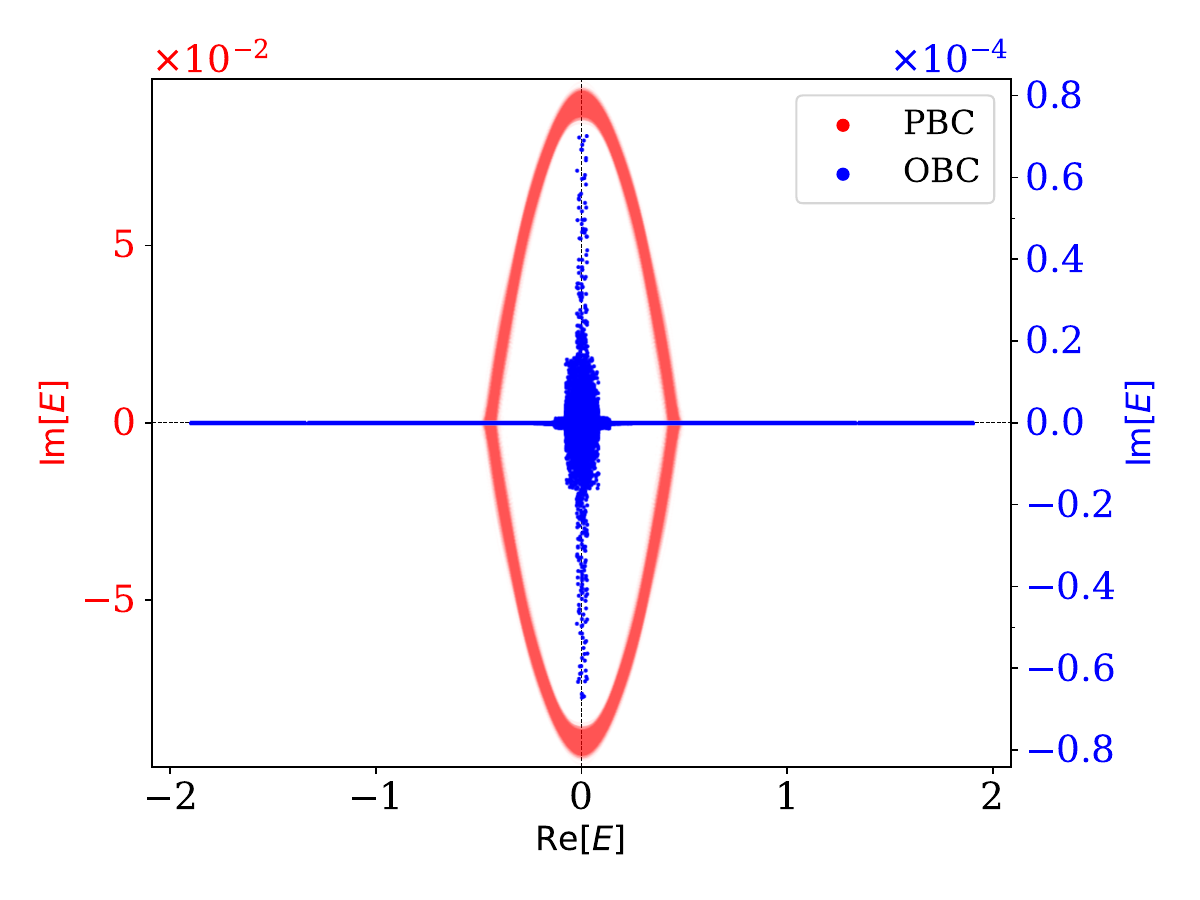}
    \caption{}\end{subfigure}\hspace{-.22cm}
    \begin{subfigure}{0.245\textwidth}
        \includegraphics[width=\linewidth]{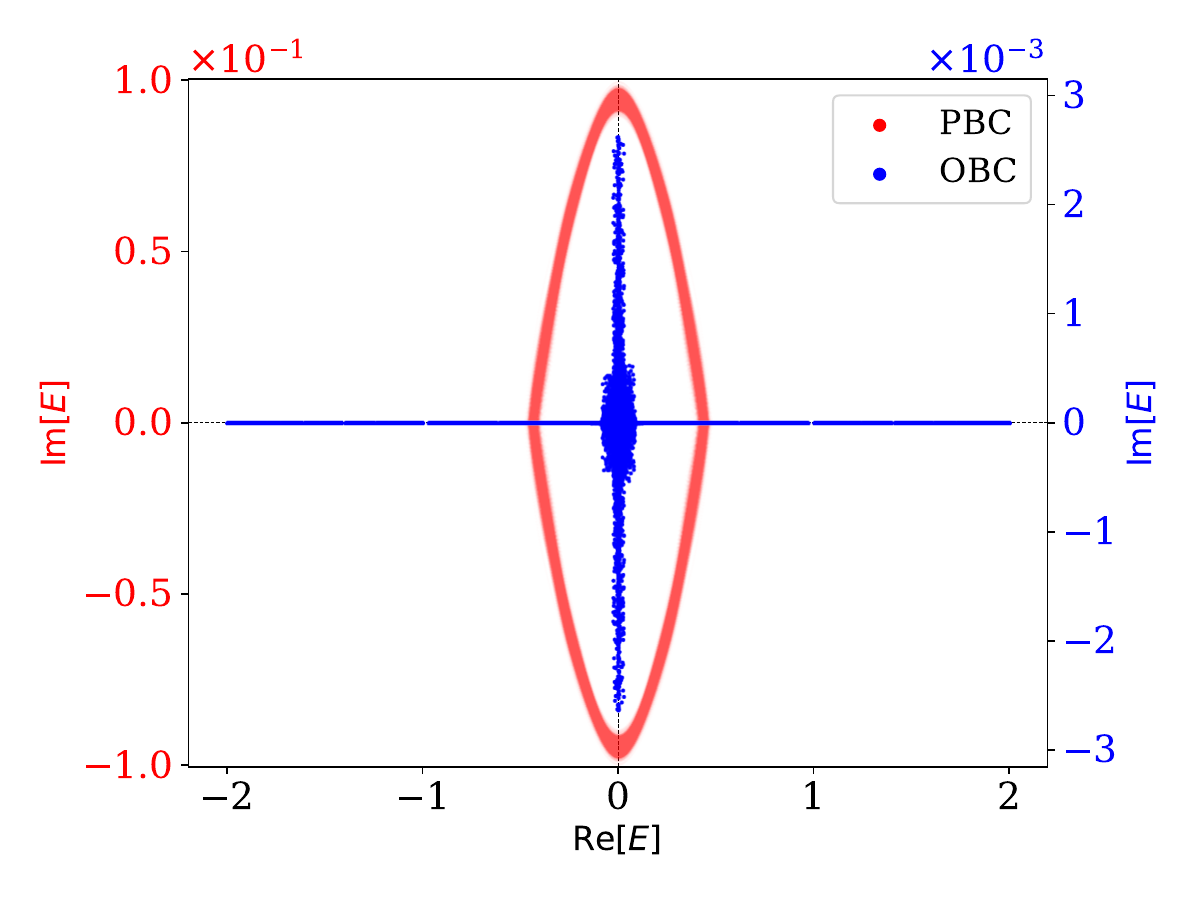}
    \caption{}
    \label{fig:spectra_symv01}\end{subfigure}
    \caption{Superimposed complex energy spectra for the symmetric impurity configuration ($h_i = h_e = 0.3$, $\gamma_e = \gamma_i = 0.1$). To properly resolve the continuous density of states, $2000$ independent random realizations are plotted together for a lattice of $N=1000$ unit cells. We fix $r=20\%$ to guarantee a macroscopic proliferation of complex eigenenergies. In alphabetic order, panels correspond to increasing hopping ratios $v/w \in \{0.65, 0.75, 0.90, 1.0\}$, mapping the emergence of the complex phase previously identified in Fig.~\ref{fig:average_2D_M_p2}. Red (blue) dots represent eigenvalues under Periodic (Open) Boundary Conditions. Notice the continuous deformation and coalescence of the disjoint PBC loops into a single boundary enclosing $E=0$ as $v/w \to 1$. Simultaneously, the imaginary energy components under OBC are drastically suppressed, illustrating the macroscopic spectral collapse characteristic of the NHSE. The geometrical behavior for $v/w > 1$ is symmetric to the $v/w < 1$ regime.}
\label{fig:spectra_sym}
\end{figure}

We now investigate the emergence of the NHSE induced by the non-hermitian defects. As $f_{\mathrm{Im}}$ remains extremely sparse at low disorder,  we set $r = 20\%$ to guarantee a robust macroscopic proliferation of complex eigenenergies which properly resolve this feature. Furthermore, because the energy spectrum of a finite disordered chain consists of discrete eigenvalues, we simulate a large system of $N = 1000$ unit cells and superimpose the complex spectra from 2000 independent random disorder realizations.

Figure~\ref{fig:spectra_sym} details the evolution of the superimposed complex energy spectra under both PBC and OBC. As the hopping ratio $v/w$ is tuned towards the maximally non-Hermitian point ($v/w=1$), the system undergoes a continuous real-complex-real spectral transition. Under PBC, the complex spectrum at $v/w = 1.0$ (Fig. \ref{fig:spectra_symv01}) forms a single macroscopic loop, a defining hallmark of non-trivial point-gap topology. As established by Gong \textit{et al.} \cite{Gong_PRX2018}, while such a loop can be continuously deformed, it can never be removed or shrunk to a single point without crossing the base point. However, as $v/w$ is detuned away from unity (e.g., towards $v/w = 0.65$ in Fig. \ref{fig:spectra_symv065}), the narrowing of the Hermitian mobility gap forces a qualitative topological change in the spectrum: the single macroscopic loop splits into two disjoint loops \cite{Gong_PRX2018} that no longer enclose the origin. This loop-splitting transition explicitly marks the point-gap topological phase transition driven by the macroscopic concentration of non-reciprocal defects. Similar loop-splitting effects have been reported for one-dimensional rings with on-site disorder \cite{Gong_PRX2018} and through spin-orbit coupling \cite{Jiang2023}, both representing extensive, non-localized effects.

In plain contrast, the behavior under OBC exhibits an extreme boundary dependence. As shown in Fig.~\ref{fig:spectra_sym}, the imaginary energy components under OBC are suppressed by several orders of magnitude (ranging from $\sim 10^{-4}$ down to $\sim 10^{-9}$) compared to the extensive macroscopic loops of the PBC case (from $\sim 10^{-1}$ to $\sim 10^{-2}$). This striking collapse is the direct mathematical manifestation of the NHSE. A non-trivial point-gap topology under PBC inherently guarantees the accumulation of an extensive number of bulk eigenstates at the open boundaries \cite{Gong_PRX2018}. In our disordered model, the non-reciprocal impurity scattering exponentially drives the hybridized wavefunctions towards the edges, effectively neutralizing the macroscopic point-gap topology and forcing the OBC spectrum back toward the real line.

\section{Asymmetric configurations: Reconfiguration of the phase boundaries}
\label{sec:asymmetric}

\begin{figure*}
    \centering
    \begin{subfigure}{0.323\textwidth}
        \centering
        \includegraphics[width=\linewidth]{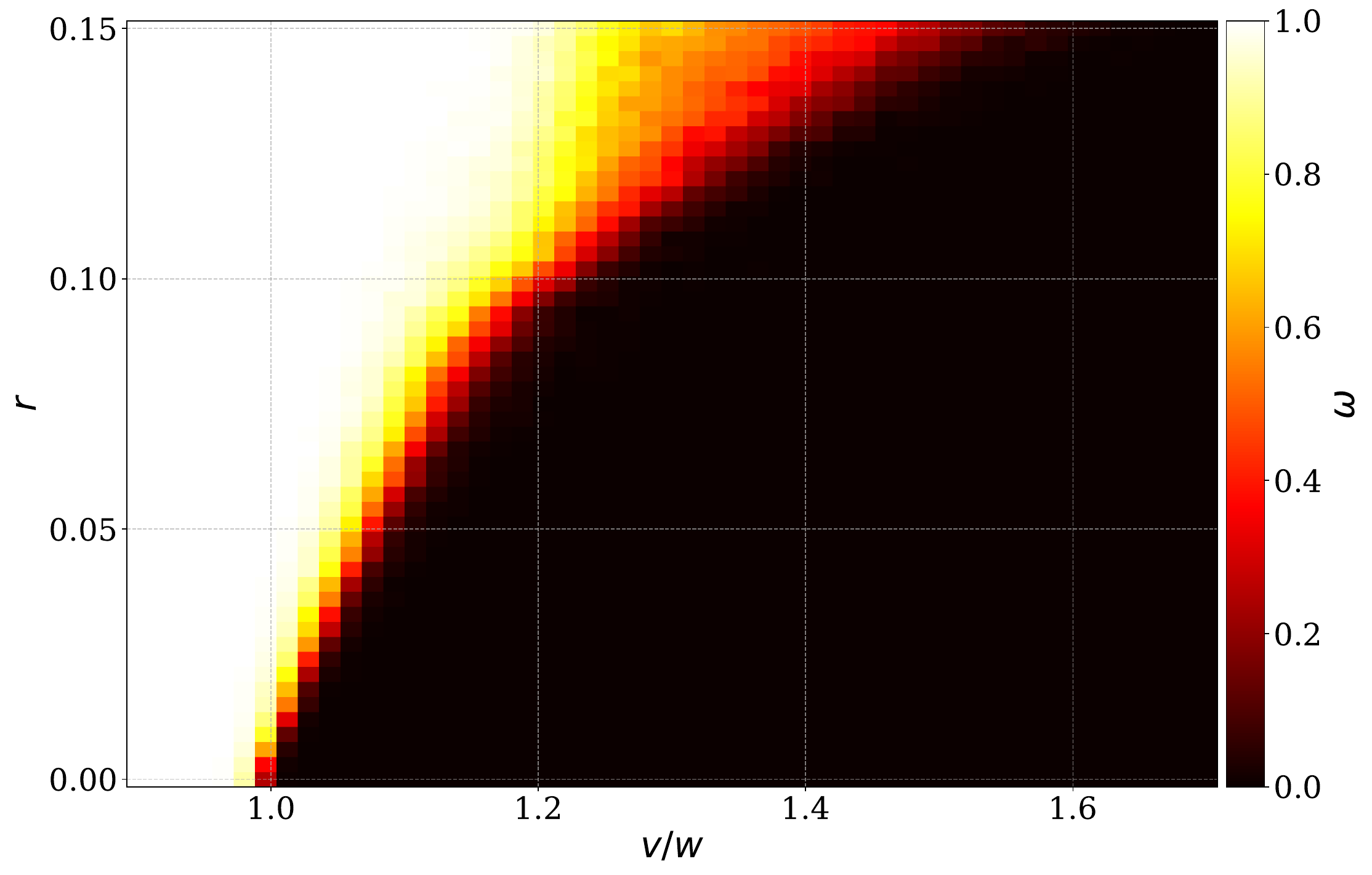}
        \caption{}
    \end{subfigure}
    \begin{subfigure}{0.323\textwidth}
        \centering
        \includegraphics[width=\linewidth]{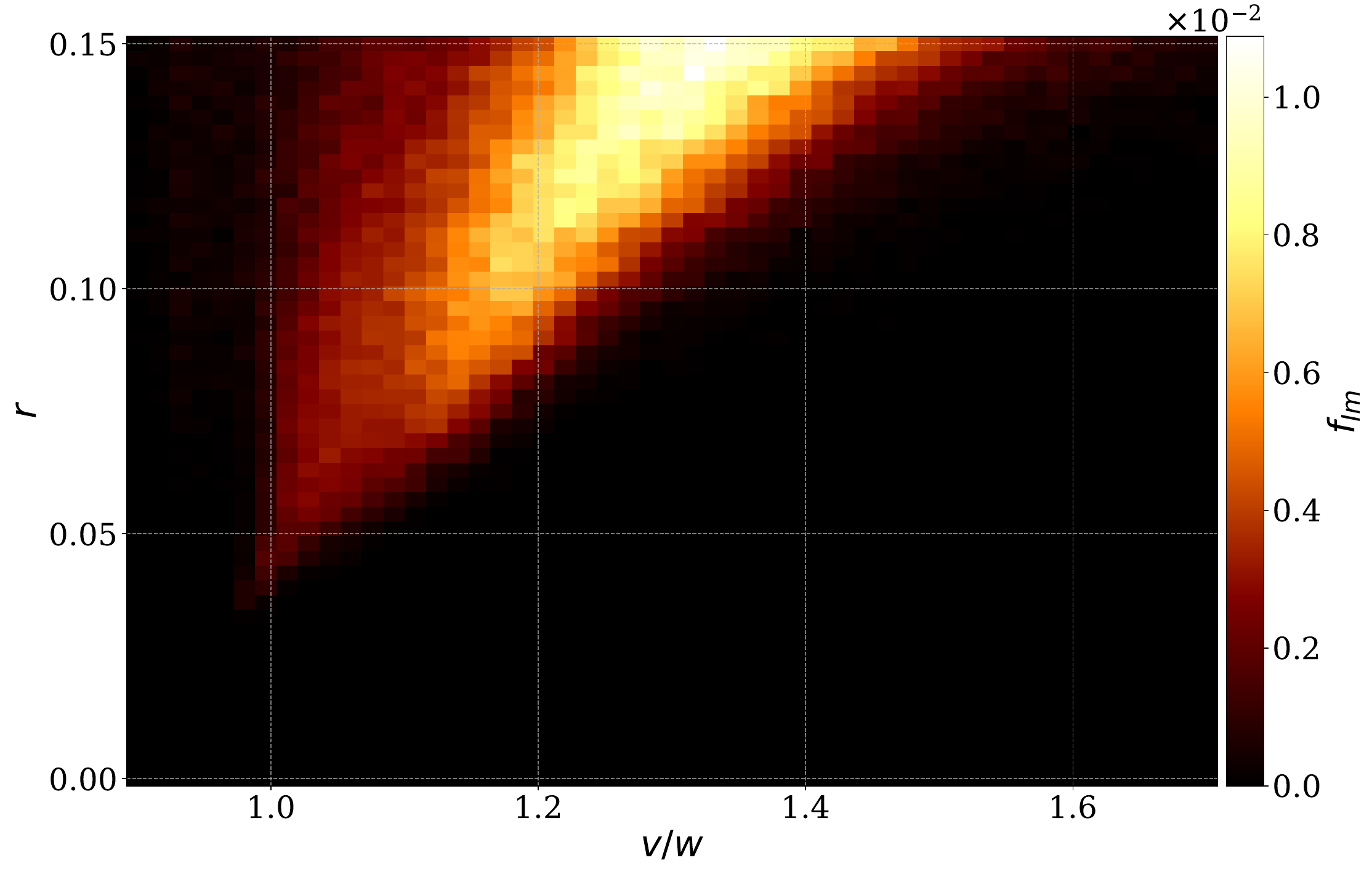}
        \caption{}
    \end{subfigure}    
\begin{subfigure}{0.323\textwidth}
        \centering
        \includegraphics[width=\linewidth]{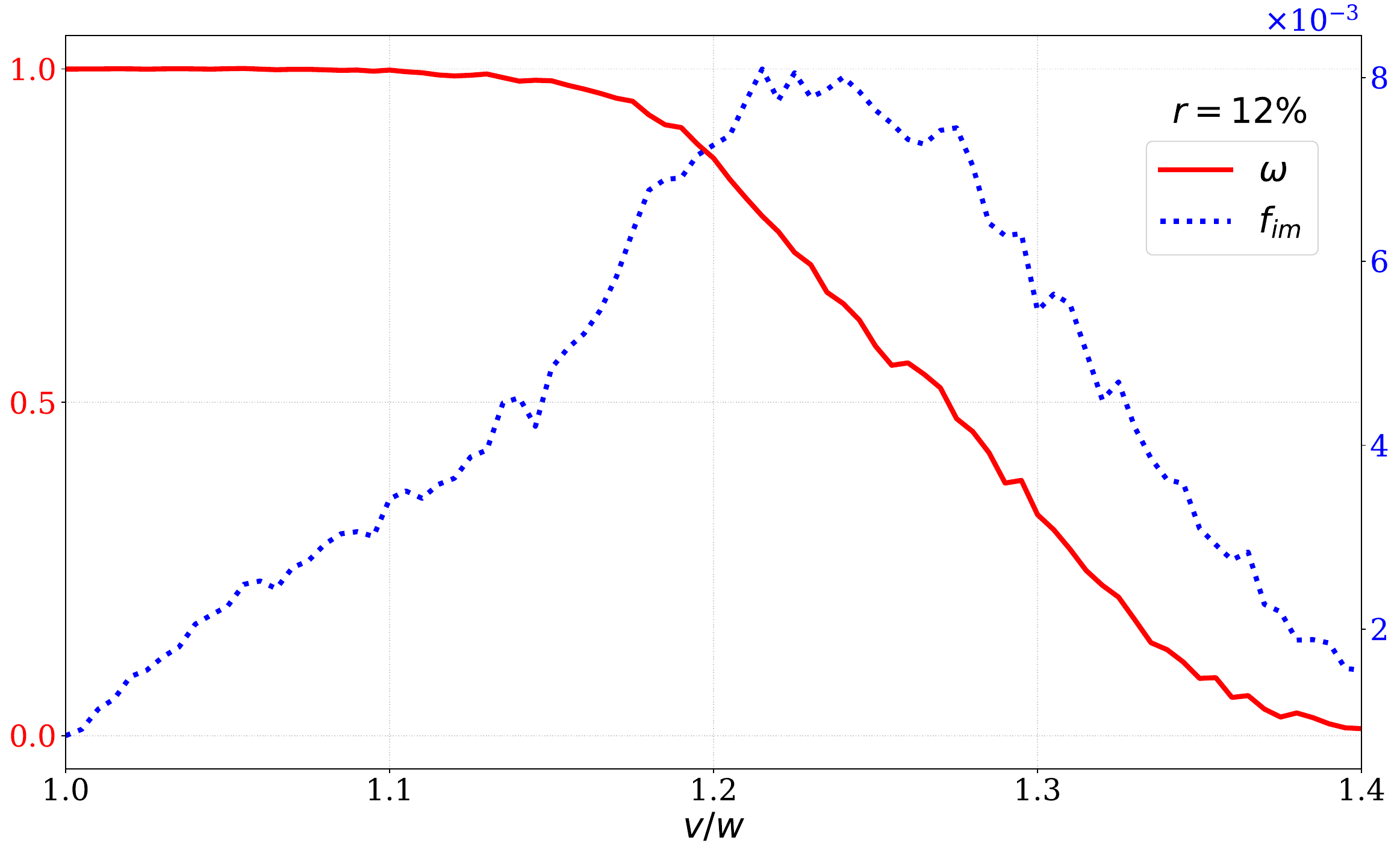}
        \caption{}
    \end{subfigure}

    \begin{subfigure}{0.323\textwidth}
        \centering
        \includegraphics[width=\linewidth]{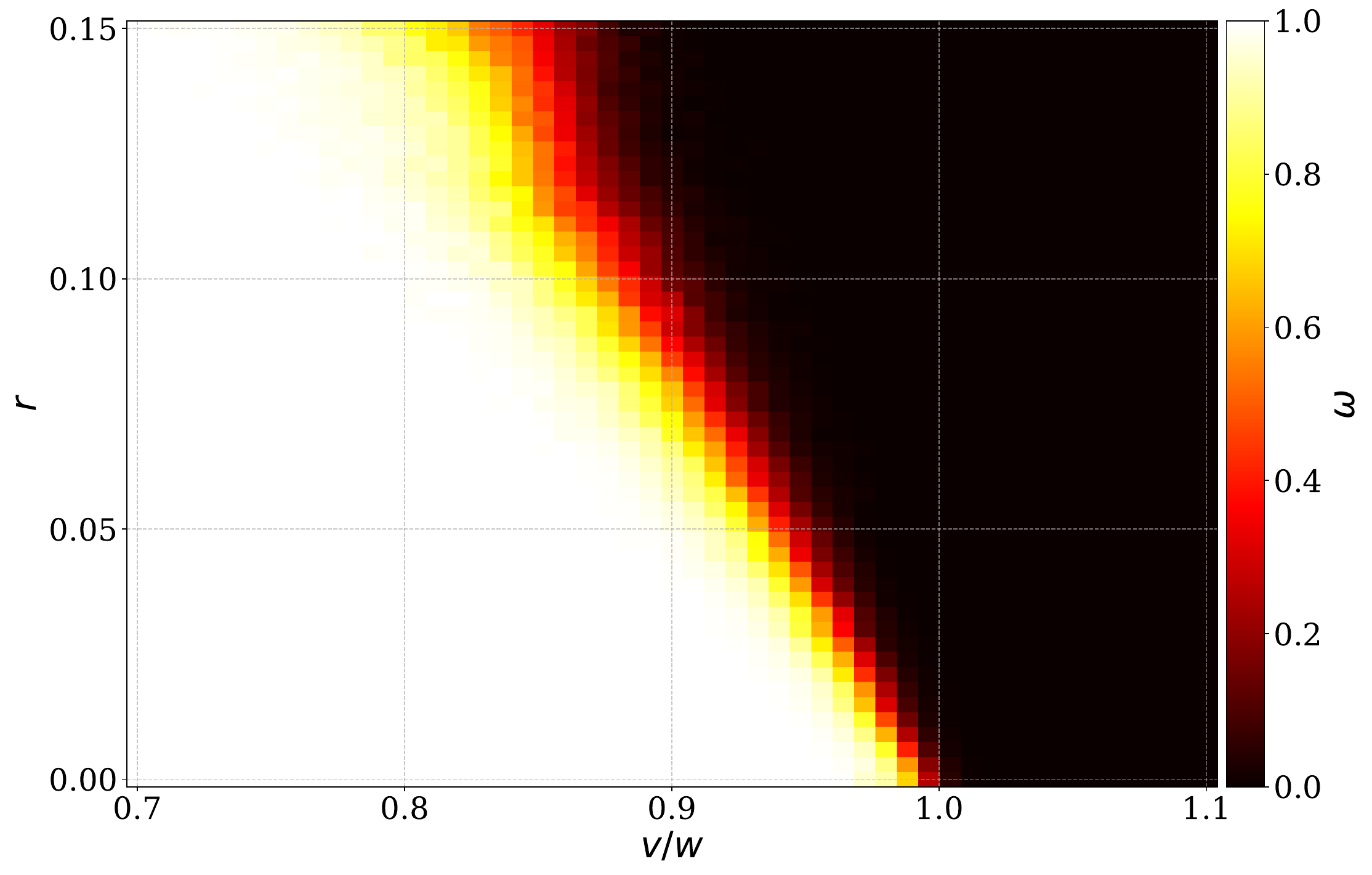}
        \caption{}
    \end{subfigure}
    \begin{subfigure}{0.323\textwidth}
        \centering
        \includegraphics[width=\linewidth]{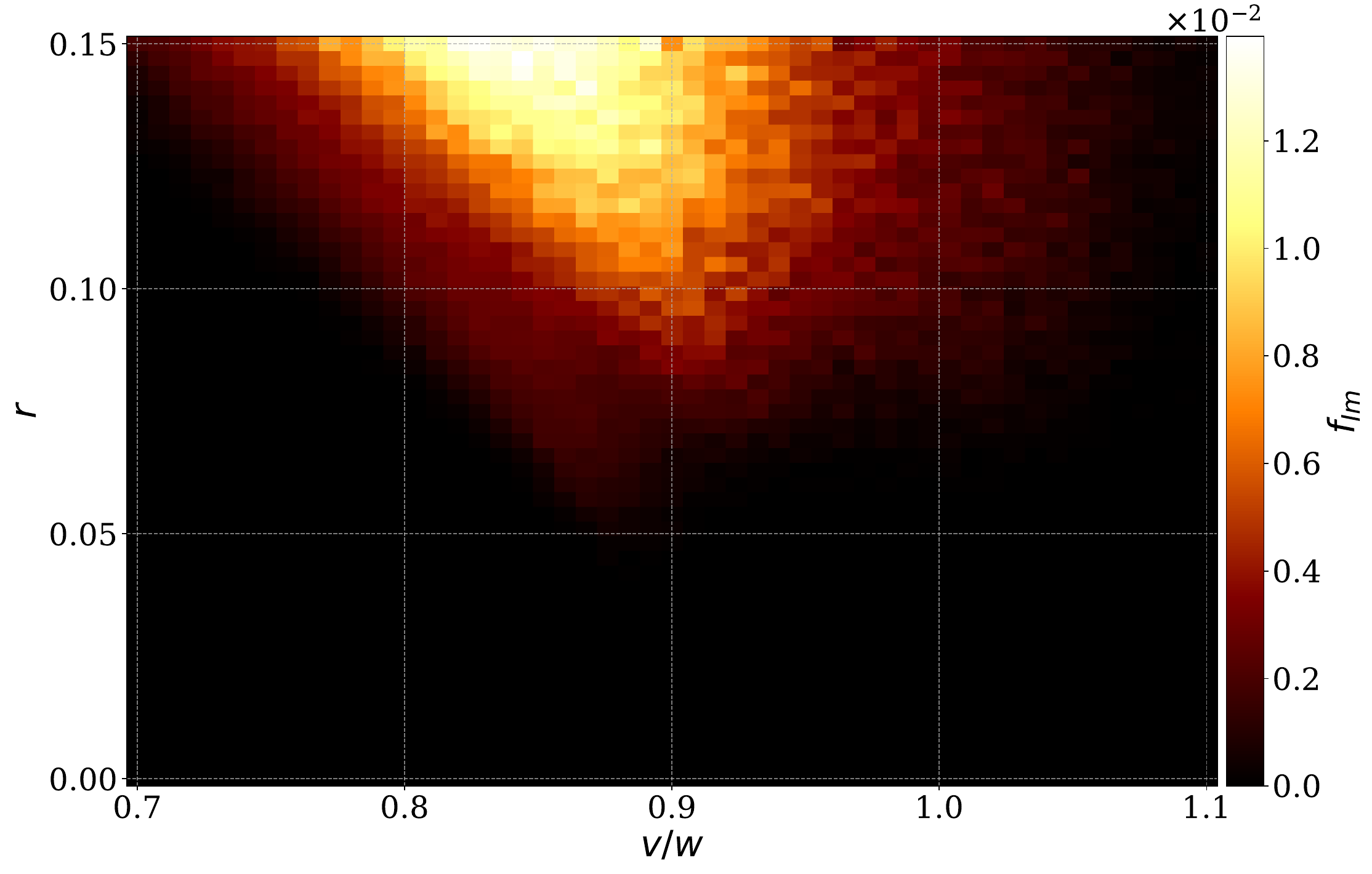}
        \caption{}
    \end{subfigure}    
\begin{subfigure}{0.323\textwidth}
        \centering
        \includegraphics[width=\linewidth]{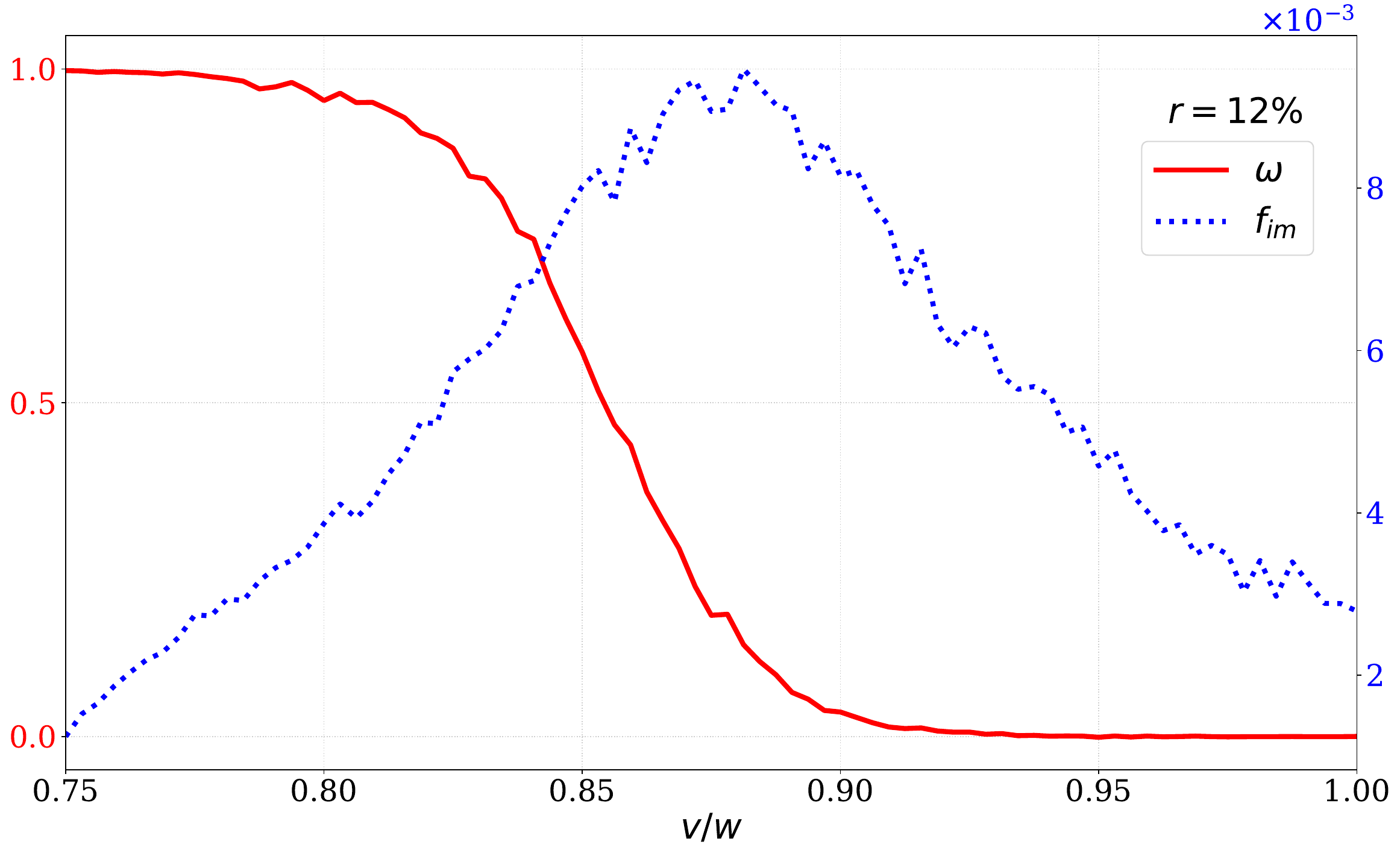}
        \caption{}
    \end{subfigure}
    \caption{Topological and spectral phase diagrams for asymmetric configurations. The left and middle columns display heat maps of the disorder-averaged real-space winding number $\omega$ ($N_r = 250$) and the complex spectrum fraction $f_{\mathrm{Im}}$ ($N_r = 750$), respectively, mapped across the $(v/w, r)$ parameter space. The rightmost column presents horizontal cross-sections at $r=12\%$ ($N_r = 1000$). The top row illustrates a configuration favoring intercell connectivity ($h_i = 0.15$, $h_e = 0.3$) alongside enhanced intracell non-reciprocity ($\gamma_i = 0.075$, $\gamma_e = 0.05$). Conversely, the bottom row depicts a configuration with suppressed intercell impurity hopping ($h_i = 0.3$, $h_e = 0.2$) and enhanced intercell non-reciprocity ($\gamma_i = 0.05$, $\gamma_e = 0.075$).}
\label{fig:phase_diagrams_asym}
\end{figure*}

We now break the spatial symmetry of the impurity parameters to study the effect of the defect's structure. Figure~\ref{fig:phase_diagrams_asym} contrasts two highly asymmetric configurations. The top row illustrates a regime where the intracell impurity hopping is suppressed ($h_i = 0.15$) relative to the intercell coupling ($h_e = 0.3$). To  physically motivate the choice of non-reciprocity parameters we bring the concept of quantum residence time (or dwell time) \cite{RevModPhys.61.917, PhysRevB.44.13582, HedayatiKh_2019}. By suppressing the intracell hopping energy, the particles are effectively more localized within the defect dimer. This prolonged residence time enhances the effective interaction with the local non-Hermitian gain-loss mechanism, naturally justifying an enhanced intracell non-reciprocity ($\gamma_i = 0.075$, $\gamma_e = 0.05$). Conversely, the bottom row depicts the complementary scenario: intercell hopping is suppressed ($h_i = 0.3$, $h_e = 0.2$), yielding an enhanced intercell non-reciprocity ($\gamma_i = 0.05$, $\gamma_e = 0.075$).

The phase diagrams for $\omega$ (left column) reveal that the configuration of the impurity hopping parameters profoundly alters the system's topological phase. A key consequence of setting $h_e > h_i$ (top row) is that the random perturbations inherently favor intercell connectivity over intracell coupling. Because robust intercell dimerization is the defining hallmark of the non-trivial SSH phase, this structural bias effectively assists the topological state. As impurities are introduced, they drag states from the bulk continuum into the energy gap to form the previously discussed impurity mini-bands. However, the asymmetric couplings shift the binding energy of these defect modes, delaying their critical hybridization with the topological edge states. This effect is a consequence of the asymmetry of impurity states energies from Fig. \ref{fig:spect_singleimp_m}: the narrow gap at $v/w < 1$ yields impurity mini-bands which push the zero-gap point into the topological trivial and smear it as $r$ increases. Consequently, the non-trivial phase is stabilized over a significantly wider parameter space, pushing the critical phase boundary deeper into the intrinsically trivial region. 

In clear contrast,  suppressing intercell hoppings ($h_i > h_e$ in bottom row) induces a structural bias that inherently favors intracell dimerization over intercell connectivity. Since the dominance of intracell hopping is the defining hallmark of the topologically trivial phase in the standard SSH model, the random defects now actively assist the trivial state. The crucial distinction in this regime lies not in the impurity concentration required to initiate the breakdown (as the invariant begins to decay at similarly low densities across both configurations) but rather in the systematic displacement of the critical point. As can be seen from the impurity state's energy asymmetry in Fig. \ref{fig:spect_singleimp_r} , the impurity mini-bands are energetically driven to hybridize prematurely with the gapless modes, causing the trivial phase to emerge earlier as $r$ increases. Consequently, the phase boundary is systematically displaced to the left, well into the intrinsically topological region ($v/w < 1$). This leftward shift dramatically compresses the parameter space where non-trivial topology can survive, representing the exact opposite of the boundary expansion observed in the intercell-dominated configuration.

This targeted displacement of the critical point is closely mirrored by the macroscopic onset of complex eigenenergies, as shown in the $f_{\mathrm{Im}}$ heat map (middle column) and the one-dimensional cross-sections at $r=12\%$ (right column). Rather than a strict point-by-point synchronization, the topological transition exhibits a broad correlation with the real-to-complex spectral breakdown, partially obscured by the residual numerical noise inherent to this extremely sparse complex regime. Furthermore, while the asymmetric configurations clearly alter the phase boundaries, their effect on the proliferation of complex states is relatively subtle. Compared to the structurally symmetric case ($h = 0.3$, $\gamma = 0.05$), where $f_{\mathrm{Im}}$ peaked at $\sim 7 \times 10^{-3}$, the asymmetric regimes reach only slightly elevated maximum values on the order of $\sim 1.0 \times 10^{-2}$ (top row) and $\sim 1.4 \times 10^{-2}$ (bottom row). This indicates that while unbalancing the impurity hoppings is highly effective at shifting the topological phase boundary, it does not drastically amplify the macroscopic susceptibility to complex eigenvalues, leaving the real spectrum remarkably stable even near the displaced critical threshold. Unlike conventional TAIs where the critical boundary shifts globally due to the variance of continuous disorder \cite{Wang_Cryst2026, Zhang_SciChP2020}, our results demonstrate that the macroscopic phase boundary can be deterministically displaced by exclusively manipulating the internal structural asymmetry of isolated non-reciprocal scatterers.

\begin{figure}[!tpb]
    \centering
    \begin{subfigure}{0.245\textwidth}
        \includegraphics[width=\linewidth]{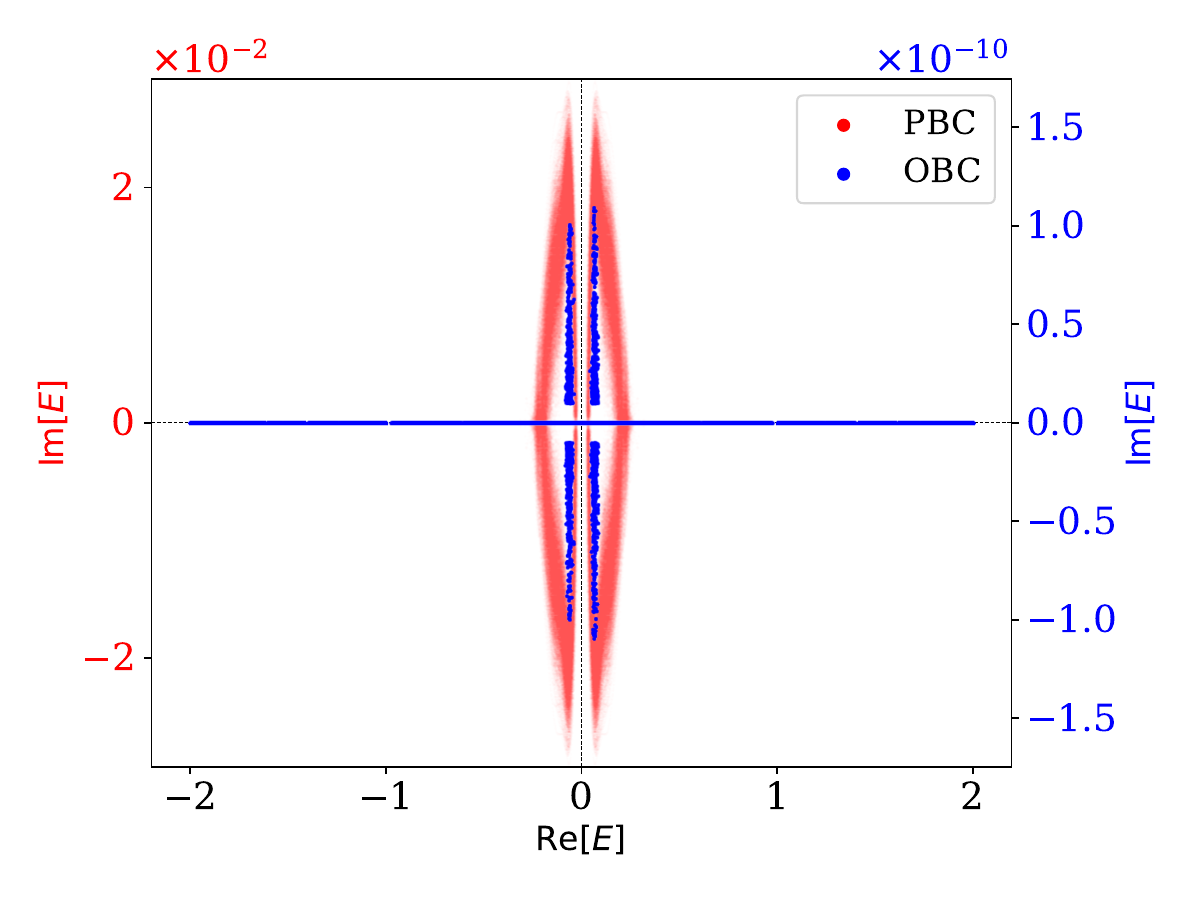}
        \caption{}
    \label{fig:spectra_asymv065} \end{subfigure}\hspace{-.22cm}
    \begin{subfigure}{0.245\textwidth}
        \includegraphics[width=\linewidth]{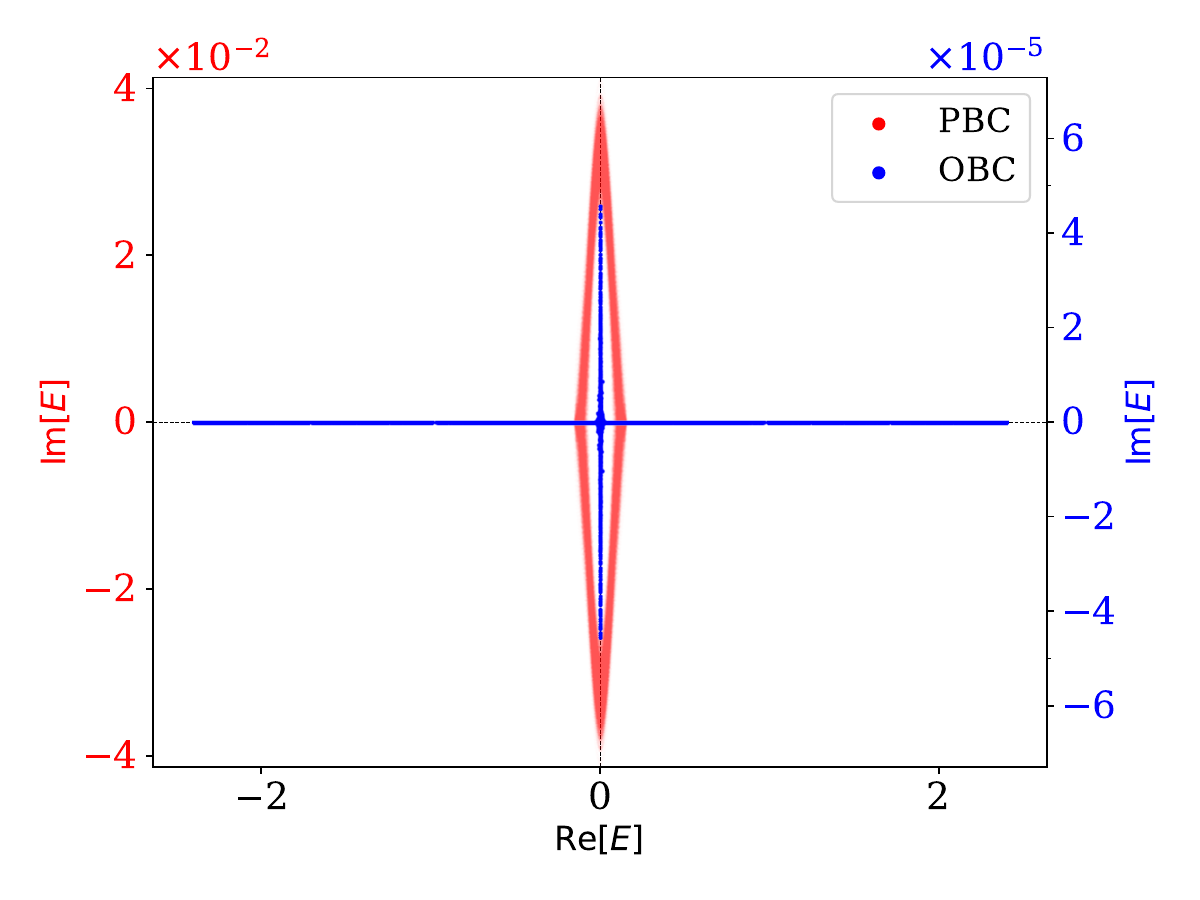}
    \caption{}\end{subfigure}
    
    \begin{subfigure}{0.245\textwidth}
        \includegraphics[width=\linewidth]{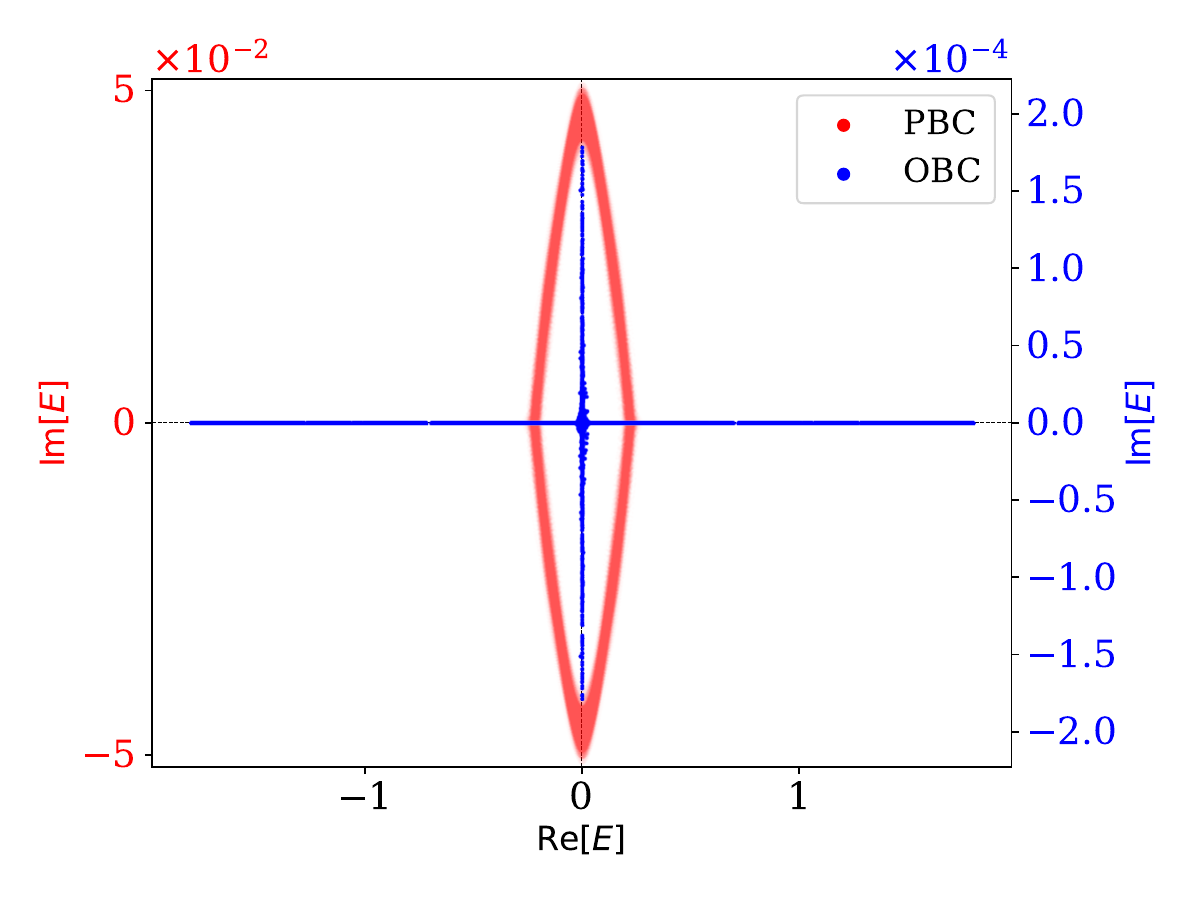}
    \caption{}\end{subfigure}\hspace{-.22cm}
    \begin{subfigure}{0.245\textwidth}
        \includegraphics[width=\linewidth]{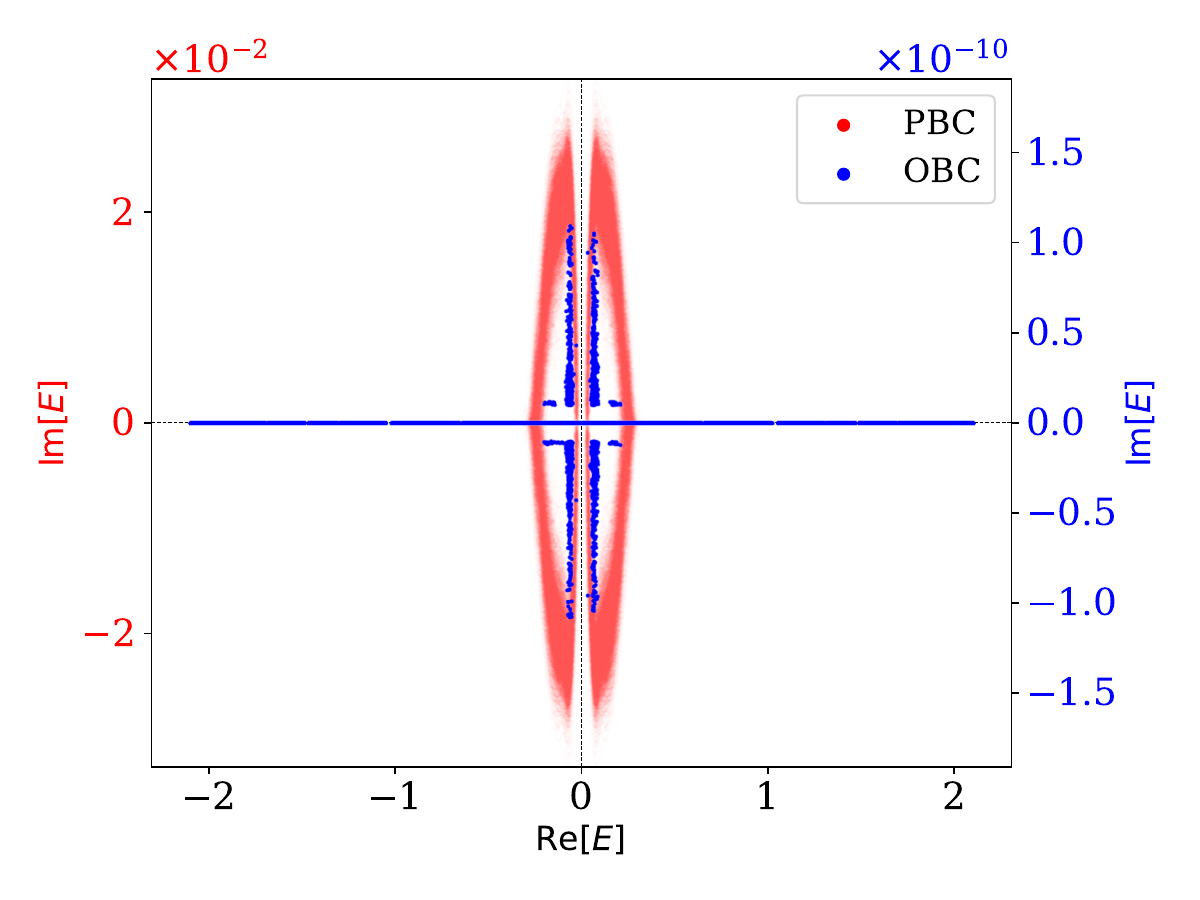}
    \caption{}
    \label{fig:spectra_asymv01}\end{subfigure}
    \caption{Superimposed complex energy spectra for two asymmetric impurity configuration.  We plot together $2000$ independent random realizations for a lattice of $N=1000$, with $r=20\%$ to guarantee a macroscopic proliferation of complex eigenenergies. Top row shows the spectra for suppressed intracell configuration ($h_i=0.15$, $h_e=0.3$, $\gamma_i = 0.075$ and $\gamma_e = 0.05$) at \textbf{(a)} $v/w=1$ and \textbf{(b)} $v/w=1.4$. Bottom row shows the spectra for suppressed intercell configuration ($h_i=0.3$, $h_e=0.2$, $\gamma_i = 0.05$ and $\gamma_e = 0.075$) at \textbf{(c)} $v/w=0.8$ and \textbf{(d)} $v/w=1.1$. Red (blue) dots represent eigenvalues under Periodic (Open) Boundary Conditions. The single/disjoint PBC loops depicted in Fig. \ref{fig:spectra_sym} are shown to be displaced according to the phase boundary discussed above, reflecting the profound impact of the asymmetry in the system.}
\label{fig:spectra_asym}
\end{figure}

We now study the NHSE induced by these asymmetric impurities. As established above, structurally asymmetric impurities systematically displace the critical phase boundary. Therefore, it becomes crucial to investigate the effect on the loop-splitting mechanism and the resulting spectral collapse seen for the symmetric configuration (Fig. \ref{fig:spectra_sym}). As in the symmetric case, we set high impurity density ($r=20\%$) and system size ($N=1000$) to achieve a clear resolution of the spectral loops.

Top and bottom rows of Fig. \ref{fig:spectra_asym} respectively shows the complex spectrum for the two asymmetric configurations given in the top and bottom rows of Fig. \ref{fig:phase_diagrams_asym}. In order to track the topological transition shown in the symmetric regime, we plot the spectra both near the displaced critical point ($v/w \approx 1.4$ at (b) for the intracell-suppressed configuration and $v/w \approx 0.8$ at (c) for the intercell-suppressed one) and the loop spliting transition ($v/w \approx 1$ at (a) for the intracell-suppressed configuration and $v/w \approx 1.1$ at (d) for the intercell-suppressed one). As clearly depicted, the macroscopic point-gap loop is identically formed but shifted to align with the new displaced critical point in each configuration. Correspondingly, the loop-splitting transition reliably manifests whenever the system is detuned away from this shifted center, accurately tracing the displaced topological boundaries. This proves that the loop-splitting mechanism and the resulting spectral collapse are not exclusive to the symmetric defect configuration. Instead, it constitutes an intrinsic effect associated with both the phase boundary reconfiguration and the smearing of the transition.

\section{Finite-Size Scaling Analysis}
\label{sec:fss}

\begin{figure*}
    \centering
        \begin{subfigure}{0.32\textwidth}
        \centering
        \includegraphics[width=\linewidth]{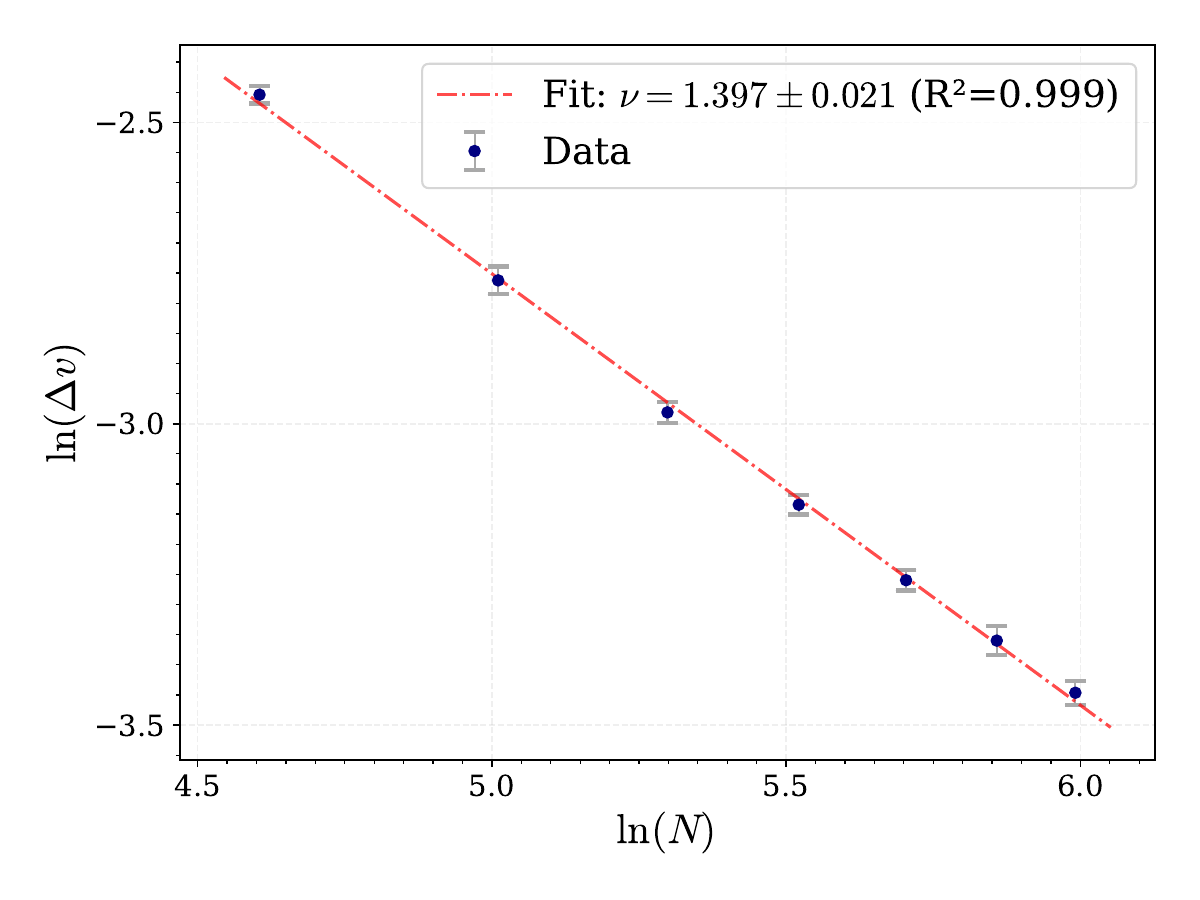}
        \caption{}
    \end{subfigure}
    \begin{subfigure}{0.32\textwidth}
        \centering
        \includegraphics[width=\linewidth]{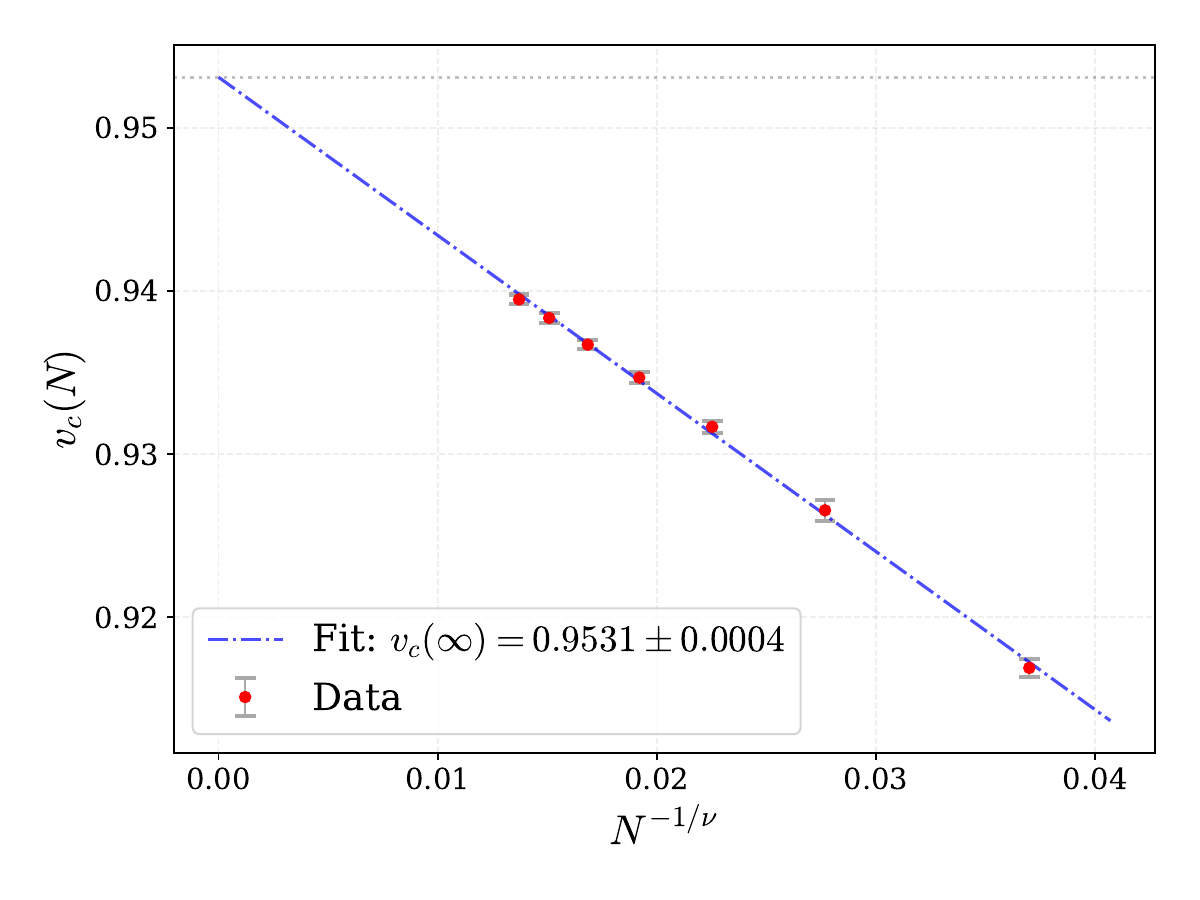}
        \caption{}
    \end{subfigure}    
\begin{subfigure}{0.32\textwidth}
        \centering
        \includegraphics[width=\linewidth]{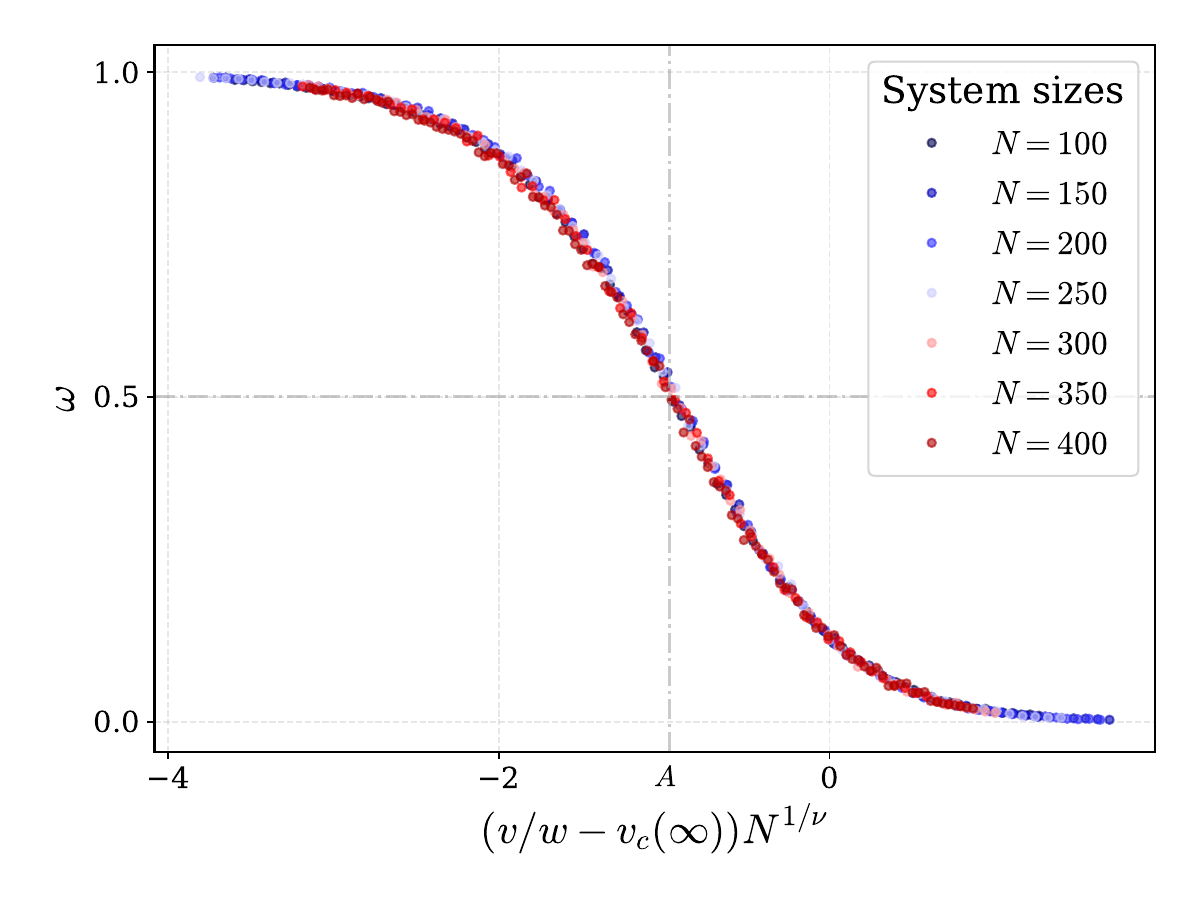}
        
        \caption{}
    \end{subfigure}
    
    \begin{subfigure}{0.32\textwidth}
        \centering
        \includegraphics[width=\linewidth]{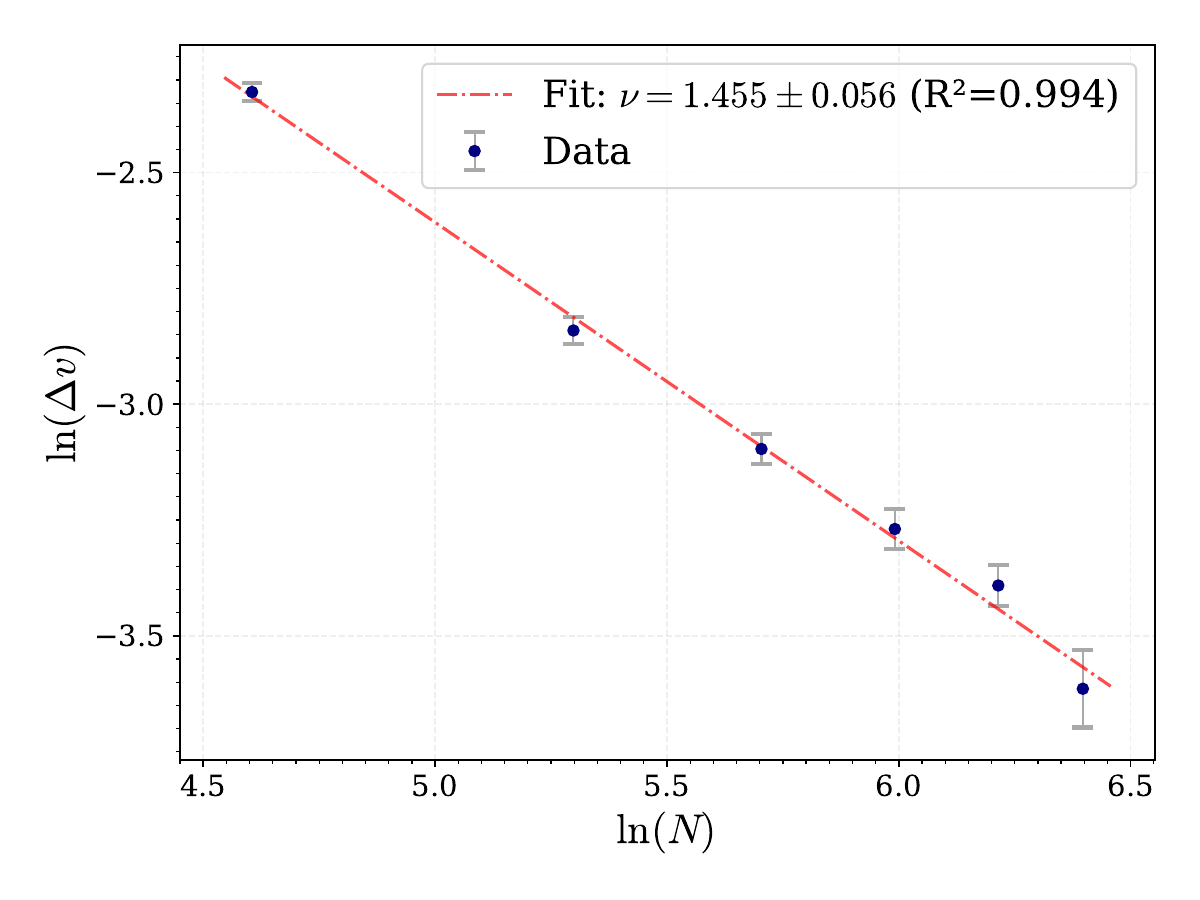}
        \caption{}
    \end{subfigure}
    \begin{subfigure}{0.32\textwidth}
        \centering
        \includegraphics[width=\linewidth]{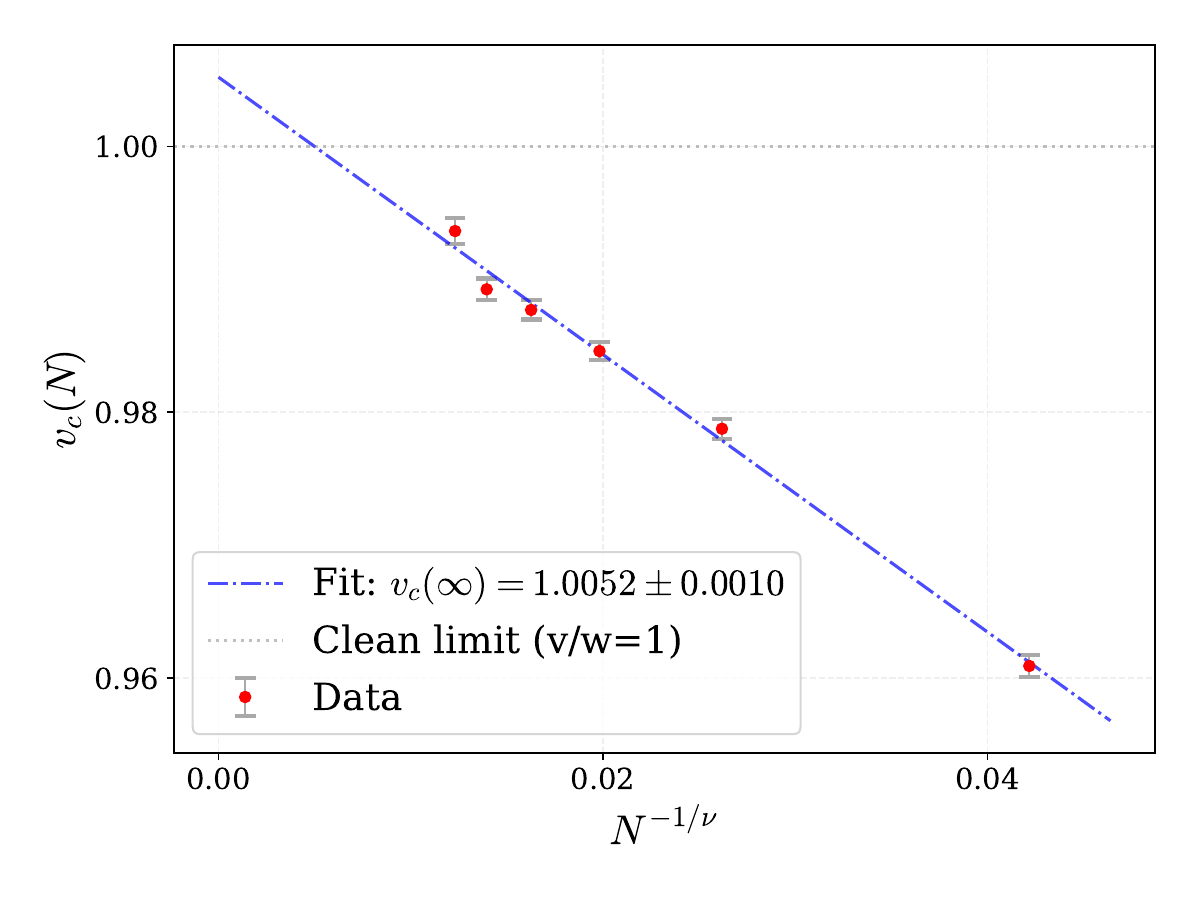}
        \caption{}
    \end{subfigure}    
\begin{subfigure}{0.32\textwidth}
        \centering
        \includegraphics[width=\linewidth]{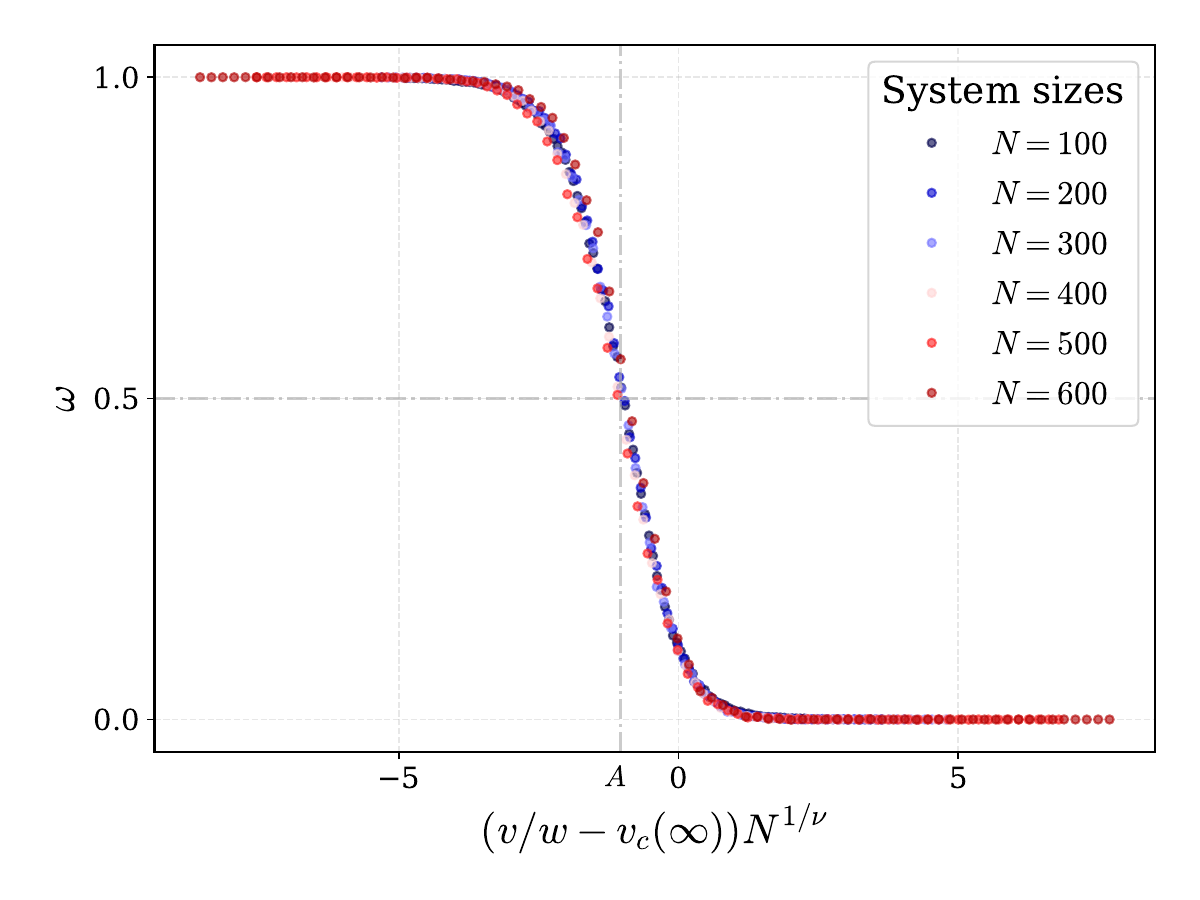}
        
        \caption{}
    \end{subfigure}

    \caption{Finite-size scaling analysis for the non-Hermitian disordered SSH chain, contrasting asymmetric and symmetric impurity configurations. The top row corresponds to an asymmetric configuration with reduced intercell hoppings ($h_i = 0.3$, $h_e = 0.2$, $\gamma_i = 0.05$, $\gamma_e = 0.075$) at $r = 5\%$. The bottom row corresponds to symmetric parameters ($h_i = h_e = 0.3$, $\gamma_i = \gamma_e = 0.05$) at $r = 10\%$. \textbf{(a, d)} Log-log scaling of the topological transition width $\Delta v$ versus system size $N$. Linear regressions (red dashed lines) yield critical exponents of $\nu \approx 1.397$ (asymmetric) and $\nu \approx 1.455$ (symmetric), confirming a disorder-induced departure from the pristine SSH universality class in both regimes. \textbf{(b, e)} Scaling of the apparent critical point $v_c(N)$ plotted against $N^{-1/\nu}$. Extrapolations to the thermodynamic limit ($N \to \infty$) give $v_c(\infty) \approx 0.9531$ for the asymmetric case and $v_c(\infty) \approx 1.0052$ for the symmetric case, rigorously confirming the macroscopic reconfiguration of the topological boundary. \textbf{(c, f)} Universal data collapse of the real-space winding number $\omega$. Plotted against the scaled variable $(v/w - v_c(\infty))N^{1/\nu}$, the macroscopic transition curves for all simulated system sizes seamlessly collapse onto a single, size-independent master curve in both configurations. Vertical dashed lines indicate the universal critical crossing coordinates.}
\label{fig:scaling_collapse}
\end{figure*}

In highly disordered non-Hermitian systems, extracting finite-size critical parameters is notoriously challenged by severe numerical instabilities (such as the biorthogonal collapse of eigenvectors near exceptional points) which manifest as unphysical spikes and heavy-tailed distributions in the winding number data. To rigorously filter these artifacts without imposing a predefined functional form, we employ a non-parametric statistical framework based on mean-averaged observables, inverse-variance weighted splines, and bootstrap resampling. A comprehensive detailing of this numerical filtering and parameter extraction protocol is provided in Appendix~\ref{app:fss_methods}.

In the thermodynamic limit ($N \to \infty$), a continuous topological phase transition is characterized by a divergent correlation length $\xi \propto |v - v_c|^{-\nu}$, where $\nu$ is the correlation length critical exponent. Physically, $\xi$ dictates the spatial decay of the zero-energy boundary states; as the system approaches criticality, these states penetrate deeper until they fully hybridize with the bulk. In a finite system, however, this macroscopic divergence is naturally cut off when the correlation length reaches the system boundary ($\xi \sim N$). Consequently, the system can no longer sustain a true mathematical singularity, and the abrupt step-like transition is smeared over a finite parameter window, $\Delta v(N)$. By substituting $\xi \to N$ and $\vert{}v - v_c\vert{} \to \Delta v(N)$ into the original divergence relation, the finite-size scaling (FSS) hypothesis \cite{Cardy_1996} dictates that this transition width must scale as:
\begin{equation}
    \Delta v(N) \propto N^{-1/\nu}.
\end{equation}

This direct link between the transition width and the topological correlation length serves as a robust diagnostic tool for identifying novel universality classes in disordered systems. In the pristine or weakly disordered regime, the correlation length exponent is analytically and numerically established to be $\nu = 1$ \cite{Aquino_PRB2023,Thatcher_PhysS2022,Chen_PRB2017}, characterizing the universality class of the standard SSH topological transition. For instance, in SSH models extended with quasiperiodic modulations and long-range hoppings, anomalous scaling regimes, including multicritical points where the exponent sharply drops below unity ($\nu \approx 0.48$) have been reported \cite{Zhang2022}. To evaluate the breakdown of this conventional topological protection and quantify the change of universality class in our non-Hermitian disordered model, we extract the critical exponent $\nu$ by performing a linear regression on the logarithmic scaling model:
\begin{equation}
    \ln(\Delta v) = -\frac{1}{\nu} \ln(N) + b.
\end{equation}

We focus our primary FSS analysis on the intercell-suppressed asymmetric configuration ($h_i=0.3, h_e=0.2, \gamma_i=0.05, \gamma_e=0.075$) at $r=5\%$, which previously exhibited a heavily displaced topological boundary. As shown in the top row of Fig.~\ref{fig:scaling_collapse}, the transition width yields a critical exponent of $\nu \approx 1.397$ [Fig.~\ref{fig:scaling_collapse}(a)]. This value marks a profound disorder-induced departure from the pristine SSH universality class ($\nu=1$). Furthermore, the apparent critical point $v_c(N)$ experiences a finite-size shift governed by this exact critical exponent. By fitting to the relation $v_c(N) = v_c(\infty) + A N^{-1/\nu}$, we extrapolate a thermodynamic threshold of $v_c(\infty) \approx 0.9531$ [Fig.~\ref{fig:scaling_collapse}(b)]. This rigorously confirms that the macroscopic displacement of the topological critical point deep into the intrinsically topological region ($v/w < 1$) is a genuine thermodynamic feature of the asymmetric non-Hermitian system, persisting unambiguously in the $N \to \infty$ limit.

While the asymmetric case highlights the deterministic tunability of the transition, we verify that this change in universality class is a general feature of macroscopic non-reciprocal disorder by evaluating the symmetric configuration ($h=0.3, \gamma=0.05$) at a moderate impurity density of $r=10\%$. Here, we similarly observe a robust linear scaling yielding a critical exponent of $\nu \approx 1.455$ [Fig.~\ref{fig:scaling_collapse}(d)], reaffirming the departure from the clean limit, alongside a subtle positive shift of the critical point to $v_c(\infty) \approx 1.0052$ [Fig.~\ref{fig:scaling_collapse}(e)].

To validate these extracted critical parameters and demonstrate that both disorder-driven transitions belong to well-defined universality classes, we perform a data collapse analysis. If the FSS hypothesis holds, the winding number curves for all simulated system sizes must collapse onto a single, size-independent universal master curve $\mathcal{F}$ when plotted against the scaled variable:
\begin{equation}
    \omega(v, N) = \mathcal{F}\big( (v - v_c(\infty)) N^{1/\nu} \big).
\end{equation}
Figures~\ref{fig:scaling_collapse}(c) and \ref{fig:scaling_collapse}(f) confirm this hypothesis for the symmetric and asymmetric configurations, respectively. In both cases, the macroscopic transition curves for all simulated system sizes seamlessly collapse onto a corresponding master curve. This successful continuous data collapse rigorously proves that the topological transition, despite its critical point being heavily displaced and its universality class altered by non-Hermitian macroscopic disorder, preserves its underlying critical nature across entirely different structural bias regimes.

Finally, the extracted critical exponent $\nu$ serves as a quantitative diagnostic tool to characterize the ultimate breakdown of topological protection. As the impurity density $r$ increases (e.g., sequentially evaluated at $r = 2.5\%$, $5\%$, and $10\%$), we observe a systematic and monotonic increase in the critical exponent $\nu$, which strictly and increasingly exceeds the pristine value of unity. This continuous growth provides direct physical evidence that the broadening of the transition is not a mere numerical artifact, but a genuine physical crossover driven by the spatial proximity of the impurities. As $r$ increases, strong defect hybridization and level repulsion induce a dense impurity mini-band that merges with the bulk states. Consequently, the effective correlation length becomes intrinsically bounded by the average inter-defect distance rather than the system size. In the highly disordered regime (e.g., $r \ge 15\%$), the transition width $\Delta v(N)$ saturates and ceases to shrink with increasing $N$. In this limit, the logarithmic slope approaches zero ($-1/\nu \to 0$), yielding an effective exponent $\nu \to \infty$. This divergence confirms the breakdown of conventional FSS and the macroscopic smearing of the non-Hermitian topological phase transition.

Before concluding this section, it is crucial to address the omission of a parallel finite-size scaling analysis for the complex spectrum fraction $f_{\mathrm{Im}}$. In the weak non-reciprocities parameter regime evaluated here, the proliferation of complex eigenenergies remains exceedingly sparse. Establishing a statistically reliable continuum for $f_{\mathrm{Im}}$ to perform a rigorous scaling analysis would inherently require drastically increasing either the system size $N$ or the macroscopic impurity concentration $r$. However, as demonstrated by the divergence of the effective critical exponent $\nu$ above, precisely these parameter adjustments trigger the macroscopic smearing of the topological boundary, thereby breaking the foundational assumptions of the finite-size scaling framework. This fundamental complementarity, where the parametric conditions required for the statistical convergence of the non-Hermitian spectral transition inherently violate the validity regime of continuous critical scaling, precludes a meaningful finite-size scaling analysis of $f_{\mathrm{Im}}$ in this specific model.

\section{Conclusion and perspectives}
\label{sec:conclusions}

In summary, we have systematically investigated the topological and spectral properties of a Hermitian Su-Schrieffer-Heeger (SSH) chain subjected to a finite concentration of randomly distributed non-Hermitian substitutional impurities. While an isolated non-reciprocal defect perfectly preserves the quantized real-space winding number, a macroscopic concentration of such defects fundamentally alters the physical landscape. The spatial proximity of random impurities induces strong wavefunction hybridization and level repulsion, broadening localized defect modes into an impurity mini-band that systematically erodes the chiral-symmetric mobility gap.

Crucially, our primary focus has been the discovery of a deterministic mechanism to actively reconfigure the topological phase boundaries. By breaking the internal structural symmetry of the non-reciprocal impurity hoppings, the thermodynamic critical point experiences a targeted macroscopic displacement. This mechanism allows for the controlled expansion or compression of the topological phase strictly via local defect asymmetry, operating independently of the continuous variance of global disorder. Finite-size scaling analysis rigorously validated this boundary displacement, extracting correlation length critical exponents ($\nu \approx 1.455$ and $\nu \approx 1.397$ for symmetric and asymmetric configurations, respectively) that confirm a macroscopic disorder-induced departure from the pristine SSH universality class.

Alongside this topological reconfiguration, we demonstrated that the disorder-driven smearing of the phase boundary is broadly correlated with a macroscopic real-to-complex spectral transition. Although a strict point-by-point synchronization is obscured by the inherent sparsity of complex eigenenergies, this spectral breakdown is geometrically characterized under periodic boundary conditions by a loop-splitting point-gap topological transition. Concurrently, under open boundary conditions, the complex spectrum undergoes a massive suppression driven by the Non-Hermitian Skin Effect, illustrating how non-reciprocal impurity scattering exponentially drives hybridized states toward the system boundaries.

Looking forward, our results establish a robust foundation for several compelling avenues of research. First, while the present finite-size scaling framework successfully captures the topological smearing at moderate disorder, achieving full statistical convergence at extremely high impurity concentrations remains an open computational challenge; future efforts utilizing advanced non-Hermitian exact diagonalization algorithms could definitively map the asymptotic behavior of the complex spectral fraction in the dense disorder limit. Second, establishing the transport signatures of these defect-induced topological breakdowns (such as evaluating the zero-bias conductance via the Landauer-Büttiker formalism or Non-Equilibrium Green's Functions) will be crucial for connecting our spectral predictions with observable experimental phenomena. Finally, extending this defect-engineered approach to other topological lattice models or higher-dimensional systems promises to unveil how isolated non-Hermitian asymmetries cooperatively interact to deterministically shift macroscopic topological protection across different universality classes.

\section*{Acknowledgments}
J.A.C. gratefully acknowledges the support of SECIHTI through the program \textit{Becas Nacionales para estudios de Posgrado}, under grant number 4018746. A.M.-R. acknowledges financial support by DGAPA-UNAM project No. IG100224, by SECIHTI project No. CBF-2025-I-1862 and by the Marcos Moshinsky Foundation. 

\section*{Data Availability}
The data that support the findings of this article are available from the authors upon reasonable request.
 
%------------------------------------------------------

\appendix

\section{Extraction of finite-size critical parameters}
\label{app:fss_methods}

To extract the critical point and the transition width without imposing a predefined functional form, we employ a non-parametric interpolation method. In highly disordered non-Hermitian systems, the evaluation of the real-space winding number $\omega$ is notoriously plagued by severe numerical instabilities. Because the calculation of $\omega$ requires the full biorthogonal basis, it is extremely vulnerable to the non-diagonalizable nature of the Hamiltonian near exceptional points, where the eigenvectors undergo a biorthogonal collapse \cite{Bergholtz_RMP2021}. This extreme sensitivity induces heavy-tailed distributions dominated by unphysical numerical spikes in $\omega$ across individual realizations\cite{Zhang_SciChP2020,Ling_PRB2021}. The severity with which the winding number computation degrades depends critically on three system parameters:
\begin{itemize}
    \item \textbf{System size ($N$):} The condition number of the eigenvalue problem grows exponentially with $N$ \cite{Feng_PRB2025}. Resolving the exponentially small splittings in larger chains quickly exhausts machine precision, severely corrupting the computed eigenvectors. Numerical instabilities proliferate at large lattice sizes.
    \item \textbf{Impurity density ($r$):} A dense concentration of non-Hermitian scatterers broadens the local hopping distribution, significantly elevating the probability of encountering highly ill-conditioned matrices.
    \item \textbf{Non-reciprocal strength ($\gamma$):} Stronger non-reciprocity pushes the system closer to exceptional points, exponentially amplifying the eigenvectors' susceptibility to numerical noise and raising the threshold for reliable diagonalization. Numerical noise increases monotonically with $\gamma$ for fixed $N$ and $r$.
\end{itemize}

Consequently, individual diagonalization attempts frequently yield values of $\omega$ far outside the quantized $0$–$1$ range. Because the winding number is a strictly quantized topological invariant, these divergent spikes do not reflect genuine physical phase transitions but rather mathematically ill-defined numerical artifacts arising exclusively from eigenvector collapse. Recognizing that these specific instabilities impose strict limitations on the accessible parameter space for $N$, $r$, and $\gamma$, we safely and systematically filter these extreme outliers from our data to perform a mathematically sound analysis.

Given these challenges, our analysis protocol relies on a rigorous statistical framework designed to ensure accuracy and reproducibility. For a fixed system size $N$ and impurity density $r$, we construct the disorder-averaged curve $\omega_N(x)$, where $x = v/w$. To prevent overfitting to residual noise while faithfully capturing the underlying transition, the data is interpolated using a cubic smoothing spline. The smoothing condition is conservatively set to the statistically optimal value for variance-weighted data. This interpolation employs statistical weights derived from the intrinsic sample-to-sample fluctuations; specifically, the weight for each point $x_i$ is taken as $w_i = 1/\sigma_i^2$, where $\sigma_i$ is the standard deviation of the winding number across the independent disorder realizations. By naturally assigning less importance to numerically noisy data points, this inverse-variance weighting ensures a robust interpolation. Crucially, we restrict the spline fit strictly to the critical transition interval (where the curve drops from $0.96$ to $0.04$). This trimming procedure effectively eliminates the influence of the flat topological and trivial tails, which would otherwise force an overfit and lead to unphysical artifacts in the transition width.

To obtain robust estimates of the statistical uncertainty, we perform a non-parametric bootstrap resampling procedure directly from the raw lists of winding numbers stored for each realization. For each of the $2000$ bootstrap iterations, we generate a synthetic averaged curve by sampling with replacement from each raw ensemble and recomputing the mean. The weighted, trimmed spline fit is then applied to this synthetic curve to extract the three roots $v_{0.9}, v_{0.5}$, and $v_{0.1}$, defined by $\omega(v_{\alpha}) = \alpha$. This process builds empirical distributions for the apparent critical point, $v_c(N) \equiv v_{0.5}$, and the transition width, $\Delta v(N) \equiv v_{0.1} - v_{0.9}$.

The final reported parameters are the means of these empirical distributions, while the uncertainties are given by the $95\%$ confidence intervals (extracted from the $2.5$th and $97.5$th percentiles), ensuring that the estimates are completely insensitive to residual outliers. We have verified the robustness of this protocol by systematically varying the spline smoothing parameter and the trimming margins; the extracted values of $v_c$ and $\Delta v$ remain stable within the bootstrap confidence intervals for all system sizes studied. This rigorous statistical framework allows us to extract physically meaningful scaling laws even in the presence of the significant numerical noise.

\bibliography{Refs.bib}

@article{Zhang2022,
doi = {10.1088/1572-9494/ac75db},
year = {2022},
month = {jul},
publisher = {IOP Publishing},
volume = {74},
number = {7},
pages = {075702},
author = {Zhang, Wei-Jie and Wu, Yi-Piao and Tang, Ling-Zhi and Zhang, Guo-Qing},
title = {Topological and dynamical phase transitions in the Su–Schrieffer–Heeger model with quasiperiodic and long-range hoppings},
journal = {Commun. Theor. Phys.}
}

@article{Prodan2011,
doi = {10.1088/1751-8113/44/11/113001},
year = {2011},
month = {feb},
publisher = {},
volume = {44},
number = {11},
pages = {113001},
author = {Prodan, Emil},
title = {Disordered topological insulators: a non-commutative geometry perspective},
journal = {J. Phys. A: Math. Theor.}
}

@article{Titum2015,
  title = {Disorder-Induced Floquet Topological Insulators},
  author = {Titum, Paraj and Lindner, Netanel H. and Rechtsman, Mikael C. and Refael, Gil},
  journal = {Phys. Rev. Lett.},
  volume = {114},
  issue = {5},
  pages = {056801},
  numpages = {5},
  year = {2015},
  month = {Feb},
  publisher = {American Physical Society},
  doi = {10.1103/PhysRevLett.114.056801}
}

@article{Zhang2021,
  title = {Localization and topological transitions in non-Hermitian quasiperiodic lattices},
  author = {Tang, Ling-Zhi and Zhang, Guo-Qing and Zhang, Ling-Feng and Zhang, Dan-Wei},
  journal = {Phys. Rev. A},
  volume = {103},
  issue = {3},
  pages = {033325},
  numpages = {9},
  year = {2021},
  month = {Mar},
  publisher = {American Physical Society},
  doi = {10.1103/PhysRevA.103.033325}
}

@article{RevModPhys.61.917,
  title = {Tunneling times: a critical review},
  author = {Hauge, E. H. and St\o{}vneng, J. A.},
  journal = {Rev. Mod. Phys.},
  volume = {61},
  issue = {4},
  pages = {917--936},
  numpages = {0},
  year = {1989},
  month = {Oct},
  publisher = {American Physical Society},
  doi = {10.1103/RevModPhys.61.917}
}

@article{HedayatiKh_2019,
doi = {10.1088/1361-648X/aaefbd},
year = {2018},
month = {dec},
publisher = {IOP Publishing},
volume = {31},
number = {3},
pages = {035302},
author = {Hedayati Kh, Hamed and Faizabadi, Edris},
title = {Transport characteristics and dwell time in a bilayer phosphorene barrier},
journal = {J. Phys. Condens. Matter}
}

@article{PhysRevB.44.13582,
  title = {Time-dependent resonant tunneling of wave packets in the tight-binding model},
  author = {Stvneng, J. A. and Hauge, E. H.},
  journal = {Phys. Rev. B},
  volume = {44},
  issue = {24},
  pages = {13582--13594},
  numpages = {0},
  year = {1991},
  month = {Dec},
  publisher = {American Physical Society},
  doi = {10.1103/PhysRevB.44.13582}
}

@article{Jiang2023,
    author = {Jiang, Wen-Cheng and Li, Jian and Li, Qing-Xu and Zhu, Jia-Ji},
    title = {The reciprocating and bipolar non-Hermitian skin effect engineered by spin–orbit coupling},
    journal = {Appl. Phys. Lett.},
    volume = {123},
    number = {20},
    pages = {201107},
    year = {2023},
    month = {11},
    issn = {0003-6951},
    doi = {10.1063/5.0174400},
}

@book{Asboth_2015,
  title={A Short Course on Topological Insulators},
  subtitle = {Band-structure topology and edge states in one and two dimensions},
  author={J\'anos K. Asb\'oth and L\'aszl\'o Oroszl\'any and Andr\'as P\'alyi},
  publisher = {Springer Cham},
  year={2016},
  DOI = {10.1007/978-3-319-25607-8},
  series ={Lecture Notes in Physics}
}

@article{Hasan_RMP2010,
  title = {Colloquium: Topological insulators},
  author = {Hasan, M. Z. and Kane, C. L.},
  journal = {Rev. Mod. Phys.},
  volume = {82},
  issue = {4},
  pages = {3045--3067},
  numpages = {0},
  year = {2010},
  month = {Nov},
  publisher = {American Physical Society},
  doi = {10.1103/RevModPhys.82.3045}
}

@article{SSH_PRL1979,
  title = {Solitons in Polyacetylene},
  author = {Su, W. P. and Schrieffer, J. R. and Heeger, A. J.},
  journal = {Phys. Rev. Lett.},
  volume = {42},
  issue = {25},
  pages = {1698--1701},
  numpages = {0},
  year = {1979},
  month = {Jun},
  publisher = {American Physical Society},
  doi = {10.1103/PhysRevLett.42.1698}
}

@article{Bomantara_JPCM2026,
doi = {10.1088/1361-648X/ae5c4e},
year = {2026},
month = {apr},
publisher = {IOP Publishing},
volume = {38},
number = {16},
pages = {163001},
author = {Bomantara, Raditya Weda},
title = {Exploring topological phases with extended Su–Schrieffer–Heeger models},
journal = {J. Phys. Condens. Matter}
}

@article{
Benalcazar_Sci2017,
author = {Wladimir A. Benalcazar  and B. Andrei Bernevig  and Taylor L. Hughes },
title = {Quantized electric multipole insulators},
journal = {Science},
volume = {357},
number = {6346},
pages = {61-66},
year = {2017},
doi = {10.1126/science.aah6442}}

@article{Liu_PRL2017,
  title = {Novel Topological Phase with a Zero Berry Curvature},
  author = {Liu, Feng and Wakabayashi, Katsunori},
  journal = {Phys. Rev. Lett.},
  volume = {118},
  issue = {7},
  pages = {076803},
  numpages = {5},
  year = {2017},
  month = {Feb},
  publisher = {American Physical Society},
  doi = {10.1103/PhysRevLett.118.076803}
}

@article{Perez_PRB2019,
  title = {Interplay between long-range hopping and disorder in topological systems},
  author = {P\'erez-Gonz\'alez, Beatriz and Bello, Miguel and G\'omez-Le\'on, \'Alvaro and Platero, Gloria},
  journal = {Phys. Rev. B},
  volume = {99},
  issue = {3},
  pages = {035146},
  numpages = {8},
  year = {2019},
  month = {Jan},
  publisher = {American Physical Society},
  doi = {10.1103/PhysRevB.99.035146}
}

@article{LEE_ChJP2022,
title = {Winding number and Zak phase in multi-band SSH models},
journal = {Chin. J. Phys.},
volume = {78},
pages = {96-110},
year = {2022},
issn = {0577-9073},
doi = {10.1016/j.cjph.2022.05.007},
author = {Chen-Shen Lee and Iao-Fai Io and Hsien-chung Kao}
}

@article{Zhang_SciChP2020,
  author = "Dan-Wei Zhang and Ling-Zhi Tang and Li-Jun Lang and Hui Yan and Shi-Liang Zhu",
  title = "Non-Hermitian topological Anderson insulators",
  journal = "Sci. China Phys. Mech. Astron.",
  year = "2020",
  volume = "63",
  number = "6",
  pages = "267062-",
  doi = "10.1007/s11433-020-1521-9"
}

@article{Li_PRL2009,
  title = {Topological Anderson Insulator},
  author = {Li, Jian and Chu, Rui-Lin and Jain, J. K. and Shen, Shun-Qing},
  journal = {Phys. Rev. Lett.},
  volume = {102},
  issue = {13},
  pages = {136806},
  numpages = {4},
  year = {2009},
  month = {Apr},
  publisher = {American Physical Society},
  doi = {10.1103/PhysRevLett.102.136806}
}

@article{Mondragon_PRL2014,
  title = {Topological Criticality in the Chiral-Symmetric AIII Class at Strong Disorder},
  author = {Mondragon-Shem, Ian and Hughes, Taylor L. and Song, Juntao and Prodan, Emil},
  journal = {Phys. Rev. Lett.},
  volume = {113},
  issue = {4},
  pages = {046802},
  numpages = {5},
  year = {2014},
  month = {Jul},
  publisher = {American Physical Society},
  doi = {10.1103/PhysRevLett.113.046802}
}

@article{Meier_Sci2018,
author = {Eric J. Meier  and Fangzhao Alex An  and Alexandre Dauphin  and Maria Maffei  and Pietro Massignan  and Taylor L. Hughes  and Bryce Gadway },
title = {Observation of the topological Anderson insulator in disordered atomic wires},
journal = {Science},
volume = {362},
number = {6417},
pages = {929-933},
year = {2018},
doi = {10.1126/science.aat3406}}

@article{Bergholtz_RMP2021,
  title = {Exceptional topology of non-Hermitian systems},
  author = {Bergholtz, Emil J. and Budich, Jan Carl and Kunst, Flore K.},
  journal = {Rev. Mod. Phys.},
  volume = {93},
  issue = {1},
  pages = {015005},
  numpages = {31},
  year = {2021},
  month = {Feb},
  publisher = {American Physical Society},
  doi = {10.1103/RevModPhys.93.015005}
}

@article{Luo_PRL2019,
  title = {Higher-Order Topological Corner States Induced by Gain and Loss},
  author = {Luo, Xi-Wang and Zhang, Chuanwei},
  journal = {Phys. Rev. Lett.},
  volume = {123},
  issue = {7},
  pages = {073601},
  numpages = {8},
  year = {2019},
  month = {Aug},
  publisher = {American Physical Society},
  doi = {10.1103/PhysRevLett.123.073601}
}

@article{Luo_APL2023,
    author = {Luo, Xi-Wang and Zhang, Chuanwei},
    title = {Photonic topological insulators induced by non-Hermitian disorders in a coupled-cavity array},
    journal = {Appl. Phys. Lett.},
    volume = {123},
    number = {8},
    pages = {081111},
    year = {2023},
    month = {08},
    issn = {0003-6951},
    doi = {10.1063/5.0153523}
}

@Article{Liu_APhyS2022,
title = {Skin effect in disordered non-Hermitian Su-Schrieffer-Heeger},
journal = {Acta Phys. Sin.},
volume = {71},
number = {22},
pages = {227402-1-227402-7},
year = {2022},
issn = {1000-3290},
doi = {10.7498/aps.71.20221151},	
author = {Liu Jia-Lin and Pang Ting-Fang and Yang Xiao-Sen and Wang Zheng-Ling}
}

@article{Zhang_AiP2022,
author = {Xiujuan Zhang and Tian Zhang and Ming-Hui Lu and Yan-Feng Chen},
title = {A review on non-Hermitian skin effect},
journal = {Adv. Phys. X},
volume = {7},
number = {1},
pages = {2109431},
year = {2022},
publisher = {Taylor \& Francis},
doi = {10.1080/23746149.2022.2109431}
}

@article{Gong_PRX2018,
  title = {Topological Phases of Non-Hermitian Systems},
  author = {Gong, Zongping and Ashida, Yuto and Kawabata, Kohei and Takasan, Kazuaki and Higashikawa, Sho and Ueda, Masahito},
  journal = {Phys. Rev. X},
  volume = {8},
  issue = {3},
  pages = {031079},
  numpages = {33},
  year = {2018},
  month = {Sep},
  publisher = {American Physical Society},
  doi = {10.1103/PhysRevX.8.031079}
}

@Article{Wang_En2025,
AUTHOR = {Wang, Junjie and Li, Fude and Cheng, Weijun},
TITLE = {Anomalous Behavior of the Non-Hermitian Topological System with an Asymmetric Coupling Impurity},
JOURNAL = {Entropy},
VOLUME = {27},
YEAR = {2025},
NUMBER = {1},
pages = {78},
PubMedID = {39851698},
ISSN = {1099-4300},
DOI = {10.3390/e27010078}
}

@article{Spring_SciPost2024,
	title = {Lack of near-sightedness principle in non-Hermitian systems},
	pages = {153},
	author = {Spring, Helene and K\"onye, Viktor and Akhmerov, Anton R. and Fulga, Ion Cosma},
	journal = {SciPost Phys.},
	volume = {17},
	year = {2024},
	publisher = {SciPost},
	doi = {10.21468/SciPostPhys.17.6.153}
}

@article{Hetenyi_PRB2025,
  title = {Localized states and skin effect around non-Hermitian impurities in tight-binding models},
  author = {Het\'enyi, Bal\'azs and D\'ora, Bal\'azs},
  journal = {Phys. Rev. B},
  volume = {112},
  issue = {7},
  pages = {075123},
  numpages = {12},
  year = {2025},
  month = {Aug},
  publisher = {American Physical Society},
  doi = {10.1103/xbj1-hfyf}
}

@article{Shi_PRB2026,
  title = {Impurity-induced topological decomposition},
  author = {Shi, Tianxing and Zhang, Chuhang and Jin, Liang and Li, Linhu},
  journal = {Phys. Rev. B},
  volume = {113},
  issue = {17},
  pages = {174205},
  numpages = {7},
  year = {2026},
  month = {May},
  publisher = {American Physical Society},
  doi = {10.1103/qq95-lnst}
}

@book{Cardy_1996,
   title =     {Scaling and Renormalization in Statistical Physics},
   author =    {John Cardy},
   publisher = {Cambridge University Press},
   isbn =      {0521499593; 9780521499590},
   year =      {1996},
   series =    {Cambridge lecture notes in physics 5}
}

@article{Ling_PRB2021,
  title = {Real-space representation of the winding number for a one-dimensional chiral-symmetric topological insulator},
  author = {Lin, Ling and Ke, Yongguan and Lee, Chaohong},
  journal = {Phys. Rev. B},
  volume = {103},
  issue = {22},
  pages = {224208},
  numpages = {12},
  year = {2021},
  month = {Jun},
  publisher = {American Physical Society},
  doi = {10.1103/PhysRevB.103.224208}
}

@Article{Wang_Cryst2026,
AUTHOR = {Wang, Xuezhi and Li, Yunlin and Liu, Yufu and Lai, Zhen and Hang, Yu and Jiang, Xunya},
TITLE = {The Study of Topological Phase Transition of One-Dimensional Disordered Su-Schrieffer-Heeger Model},
JOURNAL = {Crystals},
VOLUME = {16},
YEAR = {2026},
NUMBER = {5},
pages = {327},
ISSN = {2073-4352},
DOI = {10.3390/cryst16050327}
}

@article{Feng_PRB2025,
  title = {Numerical instability of non-Hermitian Hamiltonian evolution},
  author = {Feng, Xu and Liu, Shuo and Zhang, Shi-Xin and Chen, Shu},
  journal = {Phys. Rev. B},
  volume = {111},
  issue = {22},
  pages = {224310},
  numpages = {18},
  year = {2025},
  month = {Jun},
  publisher = {American Physical Society},
  doi = {10.1103/g1cw-tk7f}
}

@article{Aquino_PRB2023,
  title = {Critical and noncritical non-Hermitian topological phase transitions in one-dimensional chains},
  author = {Aquino, Rui and Lopes, Nei and Barci, Daniel G.},
  journal = {Phys. Rev. B},
  volume = {107},
  issue = {3},
  pages = {035424},
  numpages = {8},
  year = {2023},
  month = {Jan},
  publisher = {American Physical Society},
  doi = {10.1103/PhysRevB.107.035424}
}

@article{Chen_PRB2017,
  title = {Correlation length, universality classes, and scaling laws associated with topological phase transitions},
  author = {Chen, Wei and Legner, Markus and R\"uegg, Andreas and Sigrist, Manfred},
  journal = {Phys. Rev. B},
  volume = {95},
  issue = {7},
  pages = {075116},
  numpages = {12},
  year = {2017},
  month = {Feb},
  publisher = {American Physical Society},
  doi = {10.1103/PhysRevB.95.075116}
}

@article{Thatcher_PhysS2022,
doi = {10.1088/1402-4896/ac4ed2},
year = {2022},
month = {feb},
publisher = {IOP Publishing},
volume = {97},
number = {3},
pages = {035702},
author = {Thatcher, Luke and Fairfield, Parker and Merlo-Ramírez, Lázaro and Merlo, Juan M},
title = {Experimental observation of topological phase transitions in a mechanical 1D-SSH model},
journal = {Phys. Scr.}
}

@article{Yao_PRL2018,
  title = {Edge States and Topological Invariants of Non-Hermitian Systems},
  author = {Yao, Shunyu and Wang, Zhong},
  journal = {Phys. Rev. Lett.},
  volume = {121},
  issue = {8},
  pages = {086803},
  numpages = {8},
  year = {2018},
  month = {Aug},
  publisher = {American Physical Society},
  doi = {10.1103/PhysRevLett.121.086803}
}

@article{Khan_JLTP2026,
  title = {Single-Impurity-Induced Localization Transitions in Electronic Systems},
  volume = {222},
  pages={51},
  ISSN = {1573-7357},
  DOI = {10.1007/s10909-026-03387-7},
  number = {2},
  journal = {J. Low Temp. Phys.},
  publisher = {Springer Science and Business Media LLC},
  author = {Khan, Niaz Ali and Jan, Munsif and Shah, Muzamil and Sajid, Muhammad and Mateen, Muhammad and Ali, Mushtaq},
  year = {2026},
  month = Feb
}

@article{Wang_PRB2025,
  title = {Multiple physical quantities sensors based on non-Hermitian topological systems with an impurity},
  author = {Wang, J. J. and Zhang, Yubin and Zhang, Xiaomin and Yi, X. X.},
  journal = {Phys. Rev. Res.},
  volume = {7},
  issue = {3},
  pages = {033133},
  numpages = {11},
  year = {2025},
  month = {Aug},
  publisher = {American Physical Society},
  doi = {10.1103/h9b6-md2f}
}

@article{Pedraza2024,

title = {Impact of impurities on the topological boundaries and edge state localization in a staggered chain of atoms: SSH model and its topoelectrical circuit realization},

volume = {70},

ISSN = {0035-001X},
pages = {040501},
DOI = {10.31349/revmexfis.70.040501},

number = {4 Jul-Aug},

journal = {Rev. Mex. Fis.},

publisher = {Sociedad Mexicana de Fisica A C},

author = {P{é}rez Pedraza, Julio César and Barrios Vargas, Jos{é} Eduardo and Raya, Alfredo},

year = {2024}
}

@article{Anzurez2025,

title = {Edge state robustness in the SSH model with impurities: Topoelectric circuit simulations toward topological heterostructures},

volume = {715},

ISSN = {0921-4526},

DOI = {10.1016/j.physb.2025.417609},

journal = {Physica B: Condens. Matter},

publisher = {Elsevier BV},

author = {Anzurez, Juan and Reyes, Israel Luna and P{é}rez-Pedraza, Julio César and Raya, Alfredo and S{á}enz-Trujillo, Antonio Ulises and S{á}nchez, Roberto Tapia and Tinoco-Santill{á}n, Galileo},

year = {2025},

month = Oct,

pages = {417609}

}

@article{DaConceicao2026,

title = {Heat Capacity Anomalies and Thermal Crossovers in Finite Su-Schrieffer-Heeger Chains},
ISSN = {1402-4896},
DOI = {10.1088/1402-4896/aea5af},
journal = {Phys. Scr.},
publisher = {IOP Publishing},
author = {Da Concei{çã}o, Carlos Magno and P{é}rez-Pedraza, Julio Cesar and Raya, Alfredo and Villavicencio, Cristian},
year = {2026},
volume = {101},
number = {40},
pages = {405901}
}
\end{document}